\documentclass[journal]{IEEEtran}
\usepackage{amsmath,amsfonts}
\usepackage{algorithmic}
\usepackage{algorithm}
\usepackage{array}
\usepackage[caption=false,font=normalsize,labelfont=sf,textfont=sf]{subfig}
\usepackage{textcomp}
\usepackage{stfloats}
\usepackage{url}
\usepackage{verbatim}
\usepackage{graphicx}
\usepackage{cite}

\usepackage{bm}
\usepackage{url}

\usepackage{cite,graphicx,amsmath,amssymb}
\usepackage{fancyhdr}
\usepackage{hhline}
\usepackage{graphicx,graphics}
\usepackage{array,color}
\usepackage{amsmath}
\usepackage{stfloats}
\usepackage[flushleft]{threeparttable}
\usepackage{booktabs}

\newtheorem{proposition}{Proposition}
\newtheorem{remark}{Remark}

\newtheorem{lemma}{Lemma}

\begin{document}
\title{CRB-Rate Tradeoff for Bistatic ISAC with Gaussian Information and Deterministic Sensing Signals}

\author{Xianxin~Song,~\IEEEmembership{Member,~IEEE}, Xianghao~Yu,~\IEEEmembership{Senior Member,~IEEE}, Jie~Xu,~\IEEEmembership{Fellow,~IEEE}, and~Derrick Wing Kwan Ng,~\IEEEmembership{Fellow,~IEEE}
\thanks{Xianxin Song and Xianghao Yu are with the Department of Electrical Engineering, City University of Hong Kong, Hong Kong, China (e-mail: xianxin.song@cityu.edu.hk, alex.yu@cityu.edu.hk). Xianghao Yu is the corresponding author.}
\thanks{Jie~Xu is with the School of Science and Engineering (SSE), the Shenzhen Future Network of Intelligence Institute (FNii-Shenzhen), and the Guangdong Provincial Key Laboratory of Future Networks of Intelligence, The Chinese University of Hong Kong (Shenzhen), Guangdong  518172, China (e-mail: xujie@cuhk.edu.cn).}
\thanks{Derrick Wing Kwan Ng is with the School of Electrical Engineering and Telecommunications, University of New South Wales, Sydney, NSW 2052, Australia (e-mail: w.k.ng@unsw.edu.au).}
}

\maketitle
\begin{abstract}
In this paper, we investigate a bistatic integrated sensing and communications (ISAC) system, consisting of a base station (BS) with multiple transmit antennas, a sensing receiver with multiple receive antennas,  a single-antenna communication user (CU), and a point target to be sensed. Specifically, the BS transmits a superposition of Gaussian information and deterministic sensing signals to support ISAC. The BS aims to deliver information symbols to the CU, while the sensing receiver aims to estimate the target's direction-of-arrival (DoA) with respect to the sensing receiver by processing the echo signals reflected by the target. For the sensing receiver, we assume that only the sequences of the deterministic sensing signals and the covariance matrix of the information signals are perfectly known, whereas the specific realizations of the information signals remain unavailable. Under this setup, we first derive the corresponding Cram\'er-Rao bounds (CRBs) for DoA estimation and propose practical estimators to  accurately  estimate the target's DoA. Subsequently, we formulate the transmit beamforming design as an optimization problem aiming to minimize the CRB, subject to a minimum signal-to-interference-plus-noise ratio (SINR) requirement at the CU and a maximum transmit power constraint at the BS. When the BS employs only Gaussian information signals, the resulting beamforming optimization problem is convex, enabling the derivation of an optimal solution. In contrast, when both Gaussian information and deterministic sensing signals are transmitted, the resulting problem is non-convex and a locally optimal solution is acquired by exploiting successive convex approximation (SCA). Finally, numerical results demonstrate that the utilization of additional deterministic sensing signals is critical for sensing performance enhancement, while solely employing Gaussian information signals leads to a notable performance degradation for target sensing. It is unveiled that the proposed transmit beamforming design achieves a superior ISAC performance boundary compared with various benchmark schemes.  
\end{abstract}
\IEEEpeerreviewmaketitle
\section{Introduction}
Integrated sensing and communications (ISAC) has been envisioned as a critical use case for future sixth-generation (6G) wireless networks, supporting various emerging applications\cite{IMT2030,3GPP,9737357,9705498,9540344}. In ISAC systems, the same infrastructures and scarce spectrum resources are simultaneously exploited for effective information transmission and environmental sensing. Despite the advantages offered by this integration, significant challenges arise due to fundamental differences in their objectives and performance requirements. In particular, the criteria for signal design  and performance evaluation metrics are all distinct between these two domains. 
From the sensing perspective, radar sensing systems aim to extract target-related parameters accurately by processing received signals. Performance metrics such as the Cram\'er-Rao bounds (CRBs) for parameter estimation and  detection probabilities for target detection are widely adopted in radar sensing\cite{steven1993fundamentals_estimation,steven1993fundamentals,richards2005fundamentals,4359542,1703855,10138058}. In practice, deterministic signals are typically preferred for sensing due to their fixed waveforms and favorable autocorrelation properties\cite{steven1993fundamentals_estimation,steven1993fundamentals,richards2005fundamentals}, which facilitate accurate parameter estimation. However, utilizing deterministic sensing signals inevitably causes additional interference that adversely affects simultaneous wireless communications, thus degrading their communication performance\cite{10217169,10440056,10879807}. 

On the other hand, from the communication perspective, transceivers aim to reliably deliver  as many information symbols as possible, with the achievable data rate is typically adopted as the communication performance metric. To maximize this metric, the transmit signals must exhibit randomness. In particular, for in additive white Gaussian noise (AWGN) channels, the transmitted information signal must follow a Gaussian distribution to maximize the achievable data rate and achieve the Shannon capacity\cite{goldsmith2005wireless}. However, employing Gaussian information signals compromises sensing performance compared with utilizing deterministic sensing signals due to the lack of predetermined sequence and weaker autocorrelation properties\cite{10147248,10206462,10471902,10596930,10645253,xie2024sensing}. Therefore, effective ISAC signal design needs to meticulously balance between the  deterministic signal favored for accurate sensing and the random signally essential for maximizing communication throughput.

Motivated by the inherent tradeoff between sensing accuracy and communication throughput, there have been various existing works investigating the ISAC performance boundaries by exploiting diverse sensing performance metrics and ISAC signal models. As initial investigations, prior works in \cite{9652071,10251151,10217169} assumed a sufficiently long sensing duration, allowing the sample covariance matrix of random information signals to closely approximate the actual covariance matrix. Under these setups, the traditional closed-form estimation CRBs derived for deterministic signals were directly utilized as sensing performance metrics. Specifically, previous work \cite{9652071} considered a multi-input multi-output (MIMO) ISAC system consisting of a multi-antenna base station (BS), multiple single-antenna communication users (CUs), and a sensing target, in which the BS transmits information signals to support ISAC. The authors in \cite{9652071} formulated a beamforming optimization problem to minimize the CRB for estimating the target's direction-of-arrival (DoA) or the entire target response matrix, subject to a minimum signal-to-interference-plus-noise ratio (SINR) requirement at the CUs and a maximum transmit power constraint at the BS. Furthermore, \cite{10251151} studied a MIMO ISAC system comprising a multicasting multi-antenna BS, multiple single-antenna CUs, and multiple sensing targets. The authors in \cite{10251151} explored the tradeoff regions between the estimation CRB and the achievable multicast communication rate for scenarios where the BS transmits only information signals or a superposition of signals with both information and sensing components, respectively. Meanwhile, \cite{10217169} investigated the performance tradeoff between the estimation CRB and the communication rate in a MIMO ISAC system with a multi-antenna BS, a multi-antenna CU, and a sensing target.

In contrast to the previous studies, e.g., \cite{9652071,10251151,10217169}, directly adopting sensing metrics derived from deterministic signals, another line of work \cite{10147248,10206462,10471902,10596930,10645253,xie2024sensing} investigated random signal-enabled ISAC, specifically focusing on scenarios with limited sensing duration. For instance, the authors in \cite{10147248,10206462} adopted the Bayesian CRB \cite{10147248} and a distortion function quantifying the difference between the actual and estimated parameters \cite{10206462} as sensing performance metrics, respectively, to analyze the tradeoff between the corresponding sensing performance and ergodic communication rates. It was shown that the Bayesian CRB or distortion function is minimized if the sample covariance matrix of the ISAC signals possesses a deterministic trace. Meanwhile, if the optimization of sensing performance yields a unique solution, the optimal sample covariance matrix becomes deterministic. Besides, the authors in \cite{10596930,10645253} leveraged the ergodic linear minimum mean squared error (ELMMSE) and the ergodic least-squares error (ELSE) as metrics to characterize sensing performance for random information signals with known realizations, considering scenarios with known and unknown target impulse response (TIR) channels, respectively. Accordingly, the proposed ELMMSE and ELSE metrics were minimized by optimizing the precoding of the transmit signal for both sensing-only and ISAC scenarios. Furthermore, prior work \cite{xie2024sensing} utilized sensing mutual information (SMI) to evaluate sensing performance for random signals with known realizations, demonstrating that the SMI  achieved by random signals is always upper-bounded by that of deterministic signals. Collectively, these prior works \cite{10147248,10206462,10471902,10596930,10645253,xie2024sensing} have shown that employing deterministic signals can achieve enhanced sensing performance, albeit with the expense of communication performance degradation compared with utilizing Gaussian signals.

Remarkably, existing works \cite{10147248,10206462,10471902,10596930,10645253,xie2024sensing} considered the monostatic sensing scenario, in which the transmitter and sensing receiver are co-located. In this case, it is reasonable to assume that the realizations of random signals are perfectly known to the sensing system, owing to the convenient information exchange between the co-located transmitter and sensing receiver. In contrast to the monostatic sensing scenario, bistatic sensing represents another typical use case \cite{3GPP,10547188}, in which the transmitter and sensing receiver are not co-located. In this scenario, due to the inherent randomness of Gaussian information signals, privacy concerns for CUs, and the practical overhead and latency involved in information exchange between separate transceivers, the exact realizations of transmitted random signals may not be accessible to the sensing receiver. Estimating target parameters without specific knowledge of transmitted random signal realizations presents a greater challenges. To the best of our knowledge, characterizing the ISAC performance boundaries and developing practical estimators under conditions where Gaussian information signal realizations are unknown to sensing receiver remains an open problem, motivating this work. 
 
In this paper, we consider a bistatic ISAC system, in which a  multi-antenna BS transmits information symbols to a single-antenna CU, and a multi-antenna  sensing receiver estimates the target's DoA with respect to (w.r.t.) the sensing receiver by processing the received signal through the BS-target-receiver link. We investigate two representative signal models: one where the BS transmits only Gaussian information signals, and another where it transmits a superposition of information and dedicated sensing signals to support ISAC. For the dedicated sensing signals, their sequences are predetermined, allowing the sensing receiver to  possess perfect prior knowledge of their realizations. Whereas, for Gaussian information signals, the sensing receiver has access only to their covariance matrix, without knowledge of the specific signal realizations. Under this setup, the main contributions of this work are summarized in the following.
\begin{itemize}
	\item We derive closed-form CRB expressions for estimating the target's DoA by calculating the Fisher information matrix (FIM). Our analysis reveals that when  information signals are Gaussian and with unknown realizations, the resulting CRB is higher compared to that with deterministic sensing signals due to the indeterminacy introduced by the randomness of transmitted signals.

	\item We develop two practical estimators for estimating the target's DoA without relying on the specific realizations of transmitted information signals. When the BS only transmits Gaussian information signals, we introduce a maximum likelihood estimation (MLE) estimator to jointly estimate the target's DoA and the magnitude of channel coefficients of the BS-target-receiver link. In the scenario where the BS transmits a superposition of information and deterministic sensing signals, we propose an efficient estimator to jointly estimate the target's DoA and the complex channel coefficients. 

	\item We optimize the transmit beamforming design to minimize the estimation CRB for the sensing-only scenario. We analytically prove that utilizing Gaussian information signals leads to a higher CRB for estimating the target's DoA compared with deterministic signals, with this  performance gap diminishing as the sensing signal-to-noise ratio (SNR) increases. Meanwhile, we rigorously demonstrate that the minimum CRB with a superposition of information and deterministic sensing signals is achieved when the transmit signals are entirely deterministic. 

	\item Subsequently, we consider the more general ISAC scenario, formulating the CRB minimization problem subject to both a minimum communication requirement at the CU and a maximum transmit power constraint at the BS. For the case with only Gaussian information signals, a globally optimal beamforming design solution is obtained. For the case with a superposition of Gaussian information and deterministic sensing signals, the SINR-constrained CRB minimization problem becomes  non-convex due to the intricate non-convex CRB expression. To address this problem, we propose an iterative algorithm capitalizing on successive convex approximation (SCA) to acquire an effective solution.
	\item Finally, we present simulation results to validate the effectiveness of our proposed estimators and transmit beamforming optimization algorithms. The simulation results illustrate that the mean squared error (MSE) of the proposed estimator  converges closely to the corresponding CRB when the sensing SNR is sufficiently high, thus verifying the correctness of our CRB derivations. Meanwhile, the proposed beamforming designs  significantly outperform existing benchmark schemes, achieving a lower estimation CRB.
\end{itemize}

The remainder of this paper is organized as follows. Section~\ref{sec:system_model} describes the system model of the considered ISAC system. Sections~\ref{sec:Gaussian_Information_Signal} and \ref{sec:both_gaussian_and_deterministic} analyze the ISAC performance for scenarios where the BS transmits Gaussian information signals only and a superposition of information and deterministic sensing signals, respectively.  Finally, Section~\ref{sec:numerical_results} provides numerical results, followed by the conclusion in Section~\ref{sec:conclusion}.

\textit{Notations:} 
We denote vectors and matrices by boldface lowercase and uppercase letters, respectively. For a square matrix $\mathbf S$, $\mathrm{tr}(\mathbf S)$, $\mathbf S^{-1}$, and $\det(\mathbf S)$ represent its trace, inverse, and determinant, respectively, and $\mathbf S \succeq \mathbf{0}$ indicates its positive semi-definiteness. For any matrix $\mathbf M$, $\mathbf M^*$, $\mathbf M^{T}$, and $\mathbf M^{H}$ denote its conjugate, transpose, and conjugate transpose, respectively. The notation $\mathcal{C N}(\mathbf{0}, \mathbf{\Sigma})$ signifies a circularly symmetric complex Gaussian random vector with zero mean and covariance matrix $\mathbf \Sigma$. The spaces of $x \times y$ real and complex matrices are denoted by $\mathbb{R}^{x \times y}$ and $\mathbb{C}^{x \times y}$, respectively. For a complex number $x$, its real and imaginary parts are denoted by $\mathrm{Re}\{x\}$ and $\mathrm{Im}\{x\}$, respectively. The expectation is denoted by $\mathbb{E}[\cdot]$. The Euclidean norm of a vector is denoted by $\|\cdot\|$. The modulus of a complex number is denoted by $|\cdot|$. The operator $\mathrm{diag}(a_1, \dots, a_N)$ forms a diagonal matrix with $a_1, \dots, a_N$ on its diagonal. The vectorization operator is denoted by $\mathrm{vec}(\cdot)$. For a complex vector $\mathbf{x}$, $\mathrm{arg}(\mathbf{x})$ returns the vector of element-wise phases. The Kronecker product is denoted by $\otimes$. The imaginary unit is denoted by $\jmath = \sqrt{-1}$. For a function $\mathbf{f}(x)$, $\frac{\partial \mathbf{f}(x)}{\partial x}$ denotes the partial derivative of $\mathbf{f}(x)$ with respect to $x$.

\section{System Model and Problem Formulation}\label{sec:system_model}

\begin{figure}[t]
        \centering
        \includegraphics[width=0.5\textwidth]{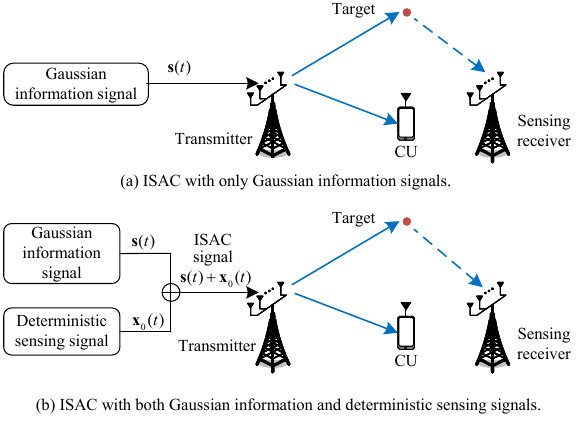}
        \caption{System model of the considered bistatic ISAC system.}
        \label{fig:system_model}
\end{figure}

As shown in Fig.~\ref{fig:system_model}, we consider a bistatic ISAC system, which consists of a BS equipped with $M_\mathrm{t}$ uniform linear array (ULA) transmit antennas, a sensing receiver equipped with $M_\mathrm{r}$ ULA receive antennas, a single-antenna CU, and a point target to be sensed.\footnote{The proposed transmit beamforming design algorithm can be readily extended to more general scenario involving multiple CUs. However, to facilitate clear exposition and insightful analysis, we focus on the single-CU scenario in this work.} In this system, the BS aims to transmit information symbols to the CU and the sensing receiver intends to estimate the target's DoA w.r.t. the sensing receiver by processing the received signals reflected by the target. 

Typically, the signal design criteria of communication and sensing are significantly distinct. For communication systems, the BS aims to reliably deliver as many as information symbols as possible. As such, the transmitted information signals usually follow a random Gaussian distribution, which maximizes the achievable data rate, attaining the Shannon capacity \cite{goldsmith2005wireless}. In contrast, target sensing systems aim to accurately extract the target's parameters by processing the received echo signals, which inherently favors deterministic signals \cite{steven1993fundamentals_estimation,steven1993fundamentals,richards2005fundamentals}. Considering these conflicting signal design requirements for sensing and communication services, we study two distinct types of transmit signal paradigms in this work: (1) the BS transmits only Gaussian information signals for ISAC, i.e., Fig.~\ref{fig:system_model}(a), and (2) the BS transmits a superposition of Gaussian information and deterministic sensing signals for ISAC, i.e., Fig.~\ref{fig:system_model}(b). In the following, we introduce the two signal transmission models.

\subsection{System Model with Only Gaussian Information Signals}
In this subsection, we present the system model when the BS exclusively transmits Gaussian information signals to simultaneously facilitate information transmission and target sensing. This scenario corresponds to a communication-centric ISAC design, where the BS directly repurposes the existing communication signals for sensing. 

Specifically, let $\mathcal T=\{1,\cdots,T\}$ denote the set of transmitted symbols, with $T$ being the number of symbols. The transmit Gaussian information signal at symbol $t\in \mathcal T$ is represented by $\mathbf s(t)\in \mathbb C^{M_\mathrm{t}\times 1}\sim \mathcal{CN}(\mathbf 0, \mathbf R_\mathrm{c})$, with zero mean and covariance matrix given by
\begin{equation}
\mathbf R_\mathrm{c} = \mathbb E\left[\mathbf s(t)\mathbf s^H(t)\right].
\end{equation}
Considering the inherent randomness of Gaussian signals, we assume that their specific realizations are {\it unknown} by the sensing receiver \cite{10547188}. Let $P$ denote the maximum transmit power at the BS. Then, the power constraint over the transmitted signals is given by
\begin{equation}
\mathbb E\left[\|\mathbf s(t)\|^2\right]=\mathrm{tr}(\mathbf R_\mathrm{c})\le P.
\end{equation}

First, we present the information transmission from the BS to the CU. Let vector $\mathbf h \in \mathbb C^{M_\mathrm{t}\times  1}$ denote the channel from the BS to the CU. Then, the received signal at the CU during symbol $t$ is given as
\begin{equation}
\begin{split}
y_{\text{c},1}(t) = \mathbf h^H \mathbf s(t) + n(t), \quad t\in \mathcal T,
\end{split}
\end{equation}
where $n(t)\sim\mathcal{CN}(0,\sigma_\mathrm{c}^2)$ denotes the AWGN at the CU receiver with noise power $\sigma_\mathrm{c}^2$. Thus, the communication SNR and the corresponding achievable communication rate at the CU are respectively given by
\begin{subequations} 
\begin{align}
\gamma_1 &=\frac{\mathbf h^H \mathbf R_\mathrm{c}\mathbf h}{\sigma_\mathrm{c}^2},\\
 R_1 &=\log_2\left(1+\gamma_1\right).
\end{align}
\end{subequations}

Next, we introduce the target sensing model. Let $\phi$ and $\theta$ denote the target's direction-of-departure (DoD) and DoA w.r.t. the BS and sensing receiver, respectively. To facilitate the subsequent CRB performance analysis, we assume that the numbers of transmit and receive antennas are both even. Without loss of generality, by choosing the midpoint of the ULA as a reference point, the steering vectors for the transmit and receive antennas towards the directions $\phi$ and $\theta$ are respectively given as
\begin{subequations}\label{eq:steering_vector}
\begin{align}\notag
&\mathbf a(\phi) \\
=& \left[e^{-\frac{\jmath\pi(M_\mathrm{t}-1)d_\mathrm{a}\sin\phi}{\lambda}}, e^{-\frac{\jmath\pi(M_\mathrm{t}-3)d_\mathrm{a}\sin\phi}{\lambda}},\cdots,e^{\frac{\jmath\pi(M_\mathrm{t}-1)d_\mathrm{a}\sin\phi}{\lambda}}\right]^T,\\\notag
&\mathbf b(\theta)\\
 =& \left[e^{-\frac{\jmath\pi(M_\mathrm{r}-1)d_\mathrm{a}\sin\theta}{\lambda}}, e^{-\frac{\jmath\pi(M_\mathrm{r}-3)d_\mathrm{a}\sin\theta}{\lambda}},\cdots,e^{\frac{\jmath\pi(M_\mathrm{r}-1)d_\mathrm{a}\sin\theta}{\lambda}}\right]^T,
\end{align}
\end{subequations}
where $d_\mathrm{a}$ denotes the spacing between two adjacent antenna elements and $\lambda$ denotes the carrier wavelength.
Then, the channel matrix of the BS-target-receiver link is given as
\begin{equation}
\mathbf H = \alpha \mathbf b(\theta) \mathbf a^T(\phi),
\end{equation}
where $\alpha=\beta \sqrt{L_1} \in\mathbb C$ denotes the channel coefficient of the BS-target-receiver link, depending on the target reflection coefficient $\beta$ and the path loss of the BS-target-receiver link $L_1$. Consequently, the received echo signal at the sensing receiver from the target at symbol $t\in\mathcal{T}$ is given as
\begin{equation}\label{eq:sensing_echo}
\begin{split}
\mathbf y_{\mathrm{s},1}(t) = \underbrace{\alpha \mathbf b(\theta) \mathbf a^T(\phi)\mathbf s(t) }_{\text{Gaussian signal}} + \underbrace{\mathbf n_\mathrm{s}(t)}_{\text{Noise}}, \quad t\in \mathcal T,
\end{split}
\end{equation}
where $\mathbf n_\mathrm{s}(t)\sim\mathcal{CN}(\mathbf 0,\mathbf I_{M_\mathrm{r}}\sigma_\mathrm{s}^2)$ is the AWGN at the sensing receiver with noise power $\sigma_\mathrm{s}^2$.

For the target sensing task, we aim to accurately determine the target's DoA w.r.t. the sensing receiver $\theta$ by processing the received echo signals in \eqref{eq:sensing_echo}. In this work, we utilize the estimation CRB, denoted as $\text{CRB}_1(\theta)$, as the sensing performance metric. Specifically, the estimation CRB provides a theoretical lower bound on the estimation error variance achievable by any unbiased estimator. The detailed CRB expression will be derived in Section~\ref{sec:CRB_Derivation}. 

Finally, with the above communication and sensing performance metrics, we aim to characterize the ISAC performance boundary achievable under various beamforming designs. In particular, let $\mathcal C_1(\mathbf R_\mathrm{c})$ denote the achievable ISAC performance region when the BS transmits only Gaussian information signals, with $\varepsilon$ and $r$ being the achievable pairs of estimation CRB and communication rate, respectively, subject to the maximum transmit power constraint at the BS. In particular, the ISAC performance region $\mathcal C_1 (\mathbf R_\mathrm{c})$ is the combination of achieved sensing and communication performance region,  defined as
\begin{equation}
\begin{split}
\mathcal C_1 (\mathbf R_\mathrm{c}) \triangleq \{(\varepsilon,r)|\varepsilon \ge\text{CRB}_1(\theta),r \le R_1,\\
\mathbf R_\mathrm{c} \succeq \mathbf 0,\mathrm{tr}(\mathbf R_\mathrm{c})\le P\}.
\end{split}
\end{equation}

\subsection{ISAC with Both Gaussian Information and Deterministic Sensing Signals}

In this subsection, we present the system model for the scenario where the BS transmits a superposition of information and dedicated sensing signals. This scenario corresponds to the case in which dedicated ISAC signals are designed for both sensing and communication services. Let  $\mathbf x_0(t)\in \mathbb C^{M_\mathrm{t}\times 1}$ denote the deterministic sensing signal transmitted by the BS at symbol $t$. Then, the sample covariance matrix of the deterministic sensing signals across the $T$ sensing symbols is
\begin{equation}
\mathbf R_\mathrm{s} =\frac{1}{T}\sum_{t=1}^T\mathbf x_0(t) \mathbf x_0^H(t).
\end{equation}
By superimposing the Gaussian information and deterministic sensing signals, the resulting ISAC signal transmitted by the BS at symbol $t$ is given as
\begin{equation}
\mathbf x(t) = \mathbf s(t) + \mathbf x_0(t), \quad t\in \mathcal T.
\end{equation}
Then, the maximum transmit power constraint at the BS becomes 
\begin{equation}
\mathrm{tr}(\mathbf R_\mathrm{c})+ \mathrm{tr}(\mathbf R_\mathrm{s})\le P.
\end{equation} 

From the communication perspective, the received signal at the CU at symbol $t$ is given as
\begin{equation}
y_{\text{c},2}(t) = \mathbf h^H \mathbf s(t) +\mathbf h^H \mathbf x_0(t)+n(t), \quad t\in \mathcal T.
\end{equation}
Note that compared with the case utilizing only Gaussian information signals for ISAC, here the CU receiver suffers from  additional interference due to the deterministic sensing signal $\mathbf x_0(t)$. Thus, the communication SINR and the corresponding achievable communication rate at the CU  are respectively given as
\begin{subequations}
\begin{align}
\gamma_2 &=\frac{\mathbf h^H \mathbf R_\mathrm{c}\mathbf h}{\mathbf h^H\mathbf R_\mathrm{s}\mathbf h + \sigma_\mathrm{c}^2},\\
R_2 &=\log_2\left(1+\gamma_2\right).
\end{align}
\end{subequations}

Next, the received echo signal at the sensing receiver, after propagation through the BS-target-receiver link at symbol $t\in\mathcal{T}$, is given as
\begin{equation}\label{eq:sensing_echo_2}
\begin{split}
\mathbf y_{\mathrm{s},2}(t) &= \underbrace{\alpha \mathbf b(\theta) \mathbf a^T(\phi)\mathbf s(t) }_{\text{Gaussian signal}}+ \underbrace{\alpha \mathbf b(\theta) \mathbf a^T(\phi) \mathbf x_0(t)}_{\text{Deterministic signal}}\\
&\quad + \underbrace{\mathbf n_\mathrm{s}(t)}_{\text{Noise}}, \quad t\in \mathcal T.
\end{split}
\end{equation}
Note that for the received echo signal $\mathbf y_{\mathrm{s},2}(t)$, both the Gaussian component $\alpha \mathbf b(\theta) \mathbf a^T(\phi)\mathbf s(t)$ and the deterministic component $\alpha \mathbf b(\theta) \mathbf a^T(\phi) \mathbf x_0(t)$ contain the information about the target. Hence, it is expected that the Gaussian and deterministic signals can be jointly utilized for a more effective target sensing. Let $\text{CRB}_2 (\theta)$ denote the CRB for estimating the target's DoA by processing the echo signal $\mathbf y_{\mathrm{s},2}(t)$ in \eqref{eq:sensing_echo_2}, which will be explicitly derived in Section~\ref{sec:CRB_Derivation}. 

Finally, let $\mathcal C_2(\mathbf R_\mathrm{c},\mathbf R_\mathrm{s})$ denote the ISAC performance region when the BS transmits both Gaussian information and deterministic sensing signals for ISAC, under the maximum transmit power constraint $P$ at the BS, which is given as
\begin{equation}
\begin{split}
\mathcal C_2(\mathbf R_\mathrm{c},\mathbf R_\mathrm{s}) \triangleq \{(\varepsilon,r)|\varepsilon \ge\text{CRB}_2(\theta),r \le R_2,\mathbf R_\mathrm{c} \succeq \mathbf 0,\\
 \mathbf R_\mathrm{s} \succeq \mathbf 0, \mathrm{tr}(\mathbf R_\mathrm{c})+ \mathrm{tr}(\mathbf R_\mathrm{s})\le P\}.
\end{split}
\end{equation} 

\begin{remark}
Utilizing only Gaussian information signals or a superposition of Gaussian information and deterministic sensing signals for ISAC each presents distinct advantages and tradeoffs. On the one hand, the scenario where the  BS transmits only Gaussian information signals can be viewed as a special case for the scenario where the BS transmits a superposition of Gaussian information and deterministic sensing signals, by directly setting  the deterministic component to be zero. Thus, it naturally follows that the ISAC performance region $\mathcal C_1(\mathbf R_\mathrm{c})$ is a subset of  $\mathcal C_2(\mathbf R_\mathrm{c},\mathbf R_\mathrm{s})$. On the other hand, although the inclusion of deterministic sensing signals offers additional degrees of freedoms (DoFs) for sensing performance enhancement, realizing these performance gains necessitates more complex hardware implementation and complicated protocol designs compared with the case with Gaussian-only signaling scenario. 
\end{remark}

\section{ISAC Performance Analysis with only Gaussian Information Signal}\label{sec:Gaussian_Information_Signal}
In this section, we investigate the CRB-rate tradeoff region when the BS transmits only Gaussian information signals. In particular, we first derive a closed-form CRB expression for estimating the target's DoA, then propose a practical MLE estimator design that utilizing only the covariance matrix of transmitted signals, and finally present optimized transmit beamforming design for both sensing-only and ISAC scenarios, respectively.

\subsection{CRB Derivation}\label{sec:CRB_Derivation}\label{sec:CRB_derivation}
First, we derive the CRB for estimating the target's DoA $\theta$. To facilitate the derivation, we stack the transmit signals, the received signals, and the noise over the whole $T$ sensing symbols as $\mathbf S= \left[\mathbf s(1),\cdots,\mathbf s(T)\right]$, $\mathbf Y_1 = \left[\mathbf y_{\mathrm{s},1}(1),\cdots,\mathbf y_{\mathrm{s},1}(T)\right]$, and $\mathbf N_\mathrm{s} = \left[\mathbf n_\mathrm{s}(1),\cdots,\mathbf n_\mathrm{s}(T)\right]$. Consequently, the received echo signal at the whole sensing time  period $T$ is stacked as
\begin{equation}\label{eq:Y}
\mathbf Y_1 = \alpha \mathbf b(\theta) \mathbf a^T(\phi) \mathbf S + \mathbf N_\mathrm{s}.
\end{equation}
By vectorizing \eqref{eq:Y} and letting $\tilde{\mathbf y}_1= \mathrm{vec}(\mathbf Y_1)$, $\tilde{\mathbf s}= \mathrm{vec}(\mathbf S)$, and $\tilde{\mathbf n}_\mathrm{s}= \mathrm{vec}(\mathbf N_\mathrm{s})$, \eqref{eq:Y} is equivalently formulated as
\begin{equation}\label{eq:echo_y_random}
\begin{split}
\tilde{\mathbf y}_1= \mathrm{vec}(\mathbf Y_1) &= \mathrm{vec}(\alpha \mathbf b(\theta) \mathbf a^T(\phi) \mathbf S) + \mathrm{vec}(\mathbf N_\mathrm{s})\\
& = \left(\mathbf I_T \otimes \alpha \mathbf b(\theta) \mathbf a^T(\phi)\right)\tilde{\mathbf s} + \tilde{\mathbf n}_\mathrm{s}\\
& = \tilde{\mathbf u}_\mathrm{c}+ \tilde{\mathbf n}_\mathrm{s},
\end{split}
\end{equation}
where $\tilde{\mathbf u}_\mathrm{c}= \left(\mathbf I_T \otimes \alpha \mathbf b(\theta) \mathbf a^T(\phi)\right)\tilde{\mathbf s}$.
For notational convenience, we express $\mathbf a(\phi)$ and $\mathbf b(\theta)$ as $\mathbf a$ and $\mathbf b$ in the following, respectively.

For the received echo signal in \eqref{eq:echo_y_random}, the sensing receiver aim to estimate  the target's DoA, i.e., $\theta$, and the complex channel coefficient, i.e., $\alpha$. 
Note that the phase of the channel coefficient $\alpha$ and the transmitted signal $\tilde{\mathbf s}$ are coupled with each other in the received component $\tilde{\mathbf u}_\mathrm{c}$. Nevertheless, the lack of specific realizations of $\tilde{\mathbf s}$ renders the phase of $\alpha$ unobservable. As an alternative, we estimate the target's DoA $\theta$ and the magnitude of channel coefficient $|\alpha|$. Let $\bm \xi=[\theta, |\alpha|]^T$ denote the vector of parameters to be estimated, and $\mathbf F \in \mathbb R^{2\times 2}$ denote the FIM for estimating $\bm \xi$, respectively. Each element of $\mathbf F$ is given by \cite{steven1993fundamentals_estimation}
\begin{equation}\label{eq:FIM}
\left[\mathbf F\right]_{(i,j)}=\mathrm{tr}\left(\mathbf R_\mathrm{n}^{-1}\frac{\partial \mathbf R_\mathrm{n}}{\partial \bm \xi_i}\mathbf R_\mathrm{n}^{-1}\frac{\partial \mathbf R_\mathrm{n}}{\partial \bm \xi_j}\right), \quad i,j\in\{1,2\},
\end{equation}
where $\mathbf R_\mathrm{n}$ is the covariance matrix of the received echo signal, i.e.,
\begin{equation}
\begin{split}
\mathbf R_\mathrm{n} &=\mathbb E\left[(\tilde{\mathbf u}_\mathrm{c}+ \tilde{\mathbf n}_\mathrm{s})(\tilde{\mathbf u}_\mathrm{c}+ \tilde{\mathbf n}_\mathrm{s})^H\right]\\
&= \mathbf I_T \otimes (|\alpha|^2\mathbf a^T \mathbf R_\mathrm{c} \mathbf a^*\mathbf b \mathbf b^H +\sigma_\mathrm{s}^2\mathbf I_M).
\end{split}
\end{equation}
Based on \eqref{eq:FIM}, the FIM for estimating $\bm \xi$ is given in the following lemma.
\begin{lemma}
Let $\dot{\mathbf b}$ denote the partial derivative of $\mathbf b$ w.r.t. $\theta$. Let $\gamma_\text{ran}(\mathbf R_\mathrm{c})= |\alpha|^2\mathbf a^T \mathbf R_\mathrm{c} \mathbf a^*\|\mathbf b\|^2/\sigma_\mathrm{s}^2$ denote the radar sensing SNR at the sensing receiver. The FIM $\mathbf F$ for estimating $\bm \xi$ is given by
\begin{equation}
\mathbf F = \begin{bmatrix}
F_{\left(\theta,\theta\right)} & F_{\left(\theta,|\alpha|\right)} \\
F_{\left(\theta,|\alpha|\right)} & F_{\left(|\alpha|,|\alpha|\right)} 
\end{bmatrix},
\end{equation}
where
\begin{subequations}
\begin{align}\label{eq:F_1}
F_{\left(\theta,\theta\right)} &= \frac{2T|\alpha|^4\left(\mathbf a^T\mathbf R_\mathrm{c}\mathbf a^*\right)^2\|\dot{\mathbf b}\|^2\|\mathbf b\|^2}{\sigma_\mathrm{s}^4(1+\gamma_\text{ran}(\mathbf R_\mathrm{c}))},\\
\label{eq:F_2}
F_{\left(\theta,|\alpha|\right)} &=0,\\
\label{eq:F_3}
F_{\left(|\alpha|,|\alpha|\right)} &=\frac{4T|\alpha|^2\left(\mathbf a^T\mathbf R_\mathrm{c}\mathbf a^*\right)^2 \|\mathbf b\|^4}{\sigma_\mathrm{s}^4(1+\gamma_\text{ran}(\mathbf R_\mathrm{c}))^2}.
\end{align}
\end{subequations}
\begin{IEEEproof}
Please refer to Appendix~\ref{sec:the_derivation_of_the_fim_mathbf_f_}.
\end{IEEEproof}
\end{lemma}

Then, the CRB for estimating the target's DoA $\theta$ is obtained as the first diagonal element of the FIM's inverse and is given as
\begin{equation}\label{eq:CRB_random_derive}
\begin{split}
&\text{CRB}_1(\theta)=\left[{\mathbf F}^{-1}\right]_{(1,1)}
= \left(F_{\left(\theta,\theta\right)} - \frac{F^2_{\left(\theta,|\alpha|\right)}}{ F_{\left(|\alpha|,|\alpha|\right)}}\right)^{-1}\\
=&  \frac{\sigma_\mathrm{s}^2}{2T|\alpha|^2\mathbf a^T\mathbf R_\mathrm{c}\mathbf a^*\|\dot{\mathbf b}\|^2}\left(1+\frac{\sigma_\mathrm{s}^2}{|\alpha|^2\mathbf a^T\mathbf R_\mathrm{c}\mathbf a^*\|\mathbf b\|^2}\right).
\end{split}
\end{equation}

Next, to gain further insight, we compare the derived CRB for estimating the target's DoA $\theta$ exploiting Gaussian information signals versus that with deterministic signals. When the transmitted signals are deterministic, the CRB for estimating $\theta$ is \cite{1703855,4359542,9652071,10596930}
\begin{equation}\label{eq:CRB_deter_derive}
\text{CRB}_\mathrm{d}(\theta) =  \frac{\sigma_\mathrm{s}^2}{2T|\alpha|^2\mathbf a^T\mathbf R_\mathrm{s}\mathbf a^*\|\dot{\mathbf b}\|^2}.
\end{equation}
Then, we have the following remark.
\begin{remark}
By comparing the CRBs in \eqref{eq:CRB_random_derive} and \eqref{eq:CRB_deter_derive}, it can be observed that  given a same beamformer design, $\mathbf R_\mathrm{c} = \mathbf R_\mathrm{s}$, the CRB for target's DoA estimation utilizing Gaussian signals suffers a scale of $\left(1+\frac{\sigma_\mathrm{s}^2}{|\alpha|^2\mathbf a^T\mathbf R_\mathrm{c}\mathbf a^*\|\mathbf b\|^2}\right)$ increment compared with that utilizing deterministic signals. Meanwhile, the performance gap between $\text{CRB}_1(\theta)$ and $\text{CRB}_\mathrm{d}(\theta)$ is expected to decrease with increasing radar sensing SNR $\gamma_\text{ran}(\mathbf R_\mathrm{c})$ and converges to zero when $\gamma_\text{ran}(\mathbf R_\mathrm{c})$ is sufficiently large. This observation will be verified in the simulation results.
\end{remark}

\vspace{-5pt}
\subsection{Practical Estimator Design}\label{sub:practical_estimator_design_1}
Then, we present a practical estimator for estimating target's DoA $\theta$ without relying on the specific realizations of transmitted signals. Based on the received echo signal model in \eqref{eq:echo_y_random}, the likelihood function for the received signal is 
\begin{equation}\label{eq:pdf_random}
p_1(\tilde{\mathbf y}_1) = \frac{\exp\left(-\tilde{\mathbf y}_1^H\mathbf R_\mathrm{n}^{-1} \tilde{\mathbf y}_1\right)}{\pi^{M_\mathrm{t}T}\det(\mathbf R_\mathrm{n})}.
\end{equation}
Next, we apply the MLE method to jointly estimate the target's DoA $\theta$ and the magnitude of channel coefficient $|\alpha|$ by maximizing the likelihood function of the received echo signal in \eqref{eq:pdf_random}, i.e.,
\begin{equation}
(\theta_\text{MLE},|\alpha|_\text{MLE}) = \arg \max_{\theta,|\alpha|}~ p_1(\tilde{\mathbf y}_1). 
\end{equation} 
To obtain the MLE value of $\theta$, we first calculate the closed-form MLE value of $|\alpha|$ with any given target's DoA $\theta$. The partial derivative of $\ln p(\tilde{\mathbf y})$ w.r.t. $|\alpha|$ is given as
\begin{equation}
\begin{split}
\frac{\partial \ln p_1(\tilde{\mathbf y}_1)}{\partial |\alpha|} &= -\mathrm{tr}\left(\mathbf R_\mathrm{n}^{-1}\frac{\partial \mathbf R_\mathrm{n}}{\partial |\alpha|}\right) + \tilde{\mathbf y}_1^H\mathbf R_\mathrm{n}^{-1} \frac{\partial \mathbf R_\mathrm{n}}{\partial |\alpha|}\mathbf R_\mathrm{n}^{-1}\tilde{\mathbf y}_1\\
&= - \frac{2T|\alpha|\mathbf a^T \mathbf R_\mathrm{c} \mathbf a^*\|\mathbf b\|^2}{\sigma_\mathrm{s}^2+|\alpha|^2\mathbf a^T \mathbf R_\mathrm{c} \mathbf a^*\|\mathbf b\|^2}\\
&\quad+ \frac{2|\alpha|\mathbf a^T \mathbf R_\mathrm{c} \mathbf a^*\sum_{t=1}^T|\mathbf b^H \mathbf y_{\mathrm{s},1}(t)|^2}{(\sigma_\mathrm{s}^2+|\alpha|^2\mathbf a^T \mathbf R_\mathrm{c} \mathbf a^*\|\mathbf b\|^2)^2}. 
\end{split}
\end{equation}
By letting $\frac{\partial \ln p_1(\tilde{\mathbf y}_1)}{\partial |\alpha|}=0$, the MLE of $|\alpha|$ under any given $\theta$ is
\begin{equation}
|\alpha|_\text{MLE}(\theta) = \sqrt{\frac{\frac{1}{T}\sum_{t=1}^T|\mathbf b^H\mathbf y_{\mathrm{s},1}(t)|^2}{\|\mathbf b\|^4\mathbf a^T\mathbf R_\mathrm{c}\mathbf a^*}-\frac{\sigma_\mathrm{s}^2}{\|\mathbf b\|^2\mathbf a^T\mathbf R_\mathrm{c}\mathbf a^*}}.
\end{equation}
Then, by substituting $|\alpha|_\text{MLE}(\theta)$ into $p_1(\tilde{\mathbf y}_1)$, we have
\begin{equation}
\begin{split}
\ln p_1(\tilde{\mathbf y}_1) =& -\frac{1}{\sigma_\mathrm{s}^2}\sum_{t=1}^T\|\mathbf y_{\mathrm{s},1}(t)\|^2-T\sigma_\mathrm{s}^2-T(M-1)\ln \sigma_\mathrm{s}^2 \\
&+ \frac{\sum_{t=1}^T|\mathbf b^H \mathbf y_{\mathrm{s},1}(t)|^2}{\|\mathbf b\|^2}\\
& - T \ln\left(\frac{\sum_{t=1}^T|\mathbf b^H \mathbf y_{\mathrm{s},1}(t)|^2}{T\|\mathbf b\|^2}\right).
\end{split}
\end{equation}
Thus, maximizing $p_1(\tilde{\mathbf y}_1)$ is equivalent to maximizing $\frac{\sum_{t=1}^T|\mathbf b^H \mathbf y_1(t)|^2}{\|\mathbf b\|^2} - T \ln\left(\frac{\sum_{t=1}^T|\mathbf b^H \mathbf y_1(t)|^2}{T\|\mathbf b\|^2}\right)$ and the MLE estimator of the target's DoA $\theta$ is
\begin{equation}\label{eq:MLE_theta}
\begin{split}
\theta_\text{MLE} = \arg \max_{\theta}~&\frac{\sum_{t=1}^T|\mathbf b^H \mathbf y_{\mathrm{s},1}(t)|^2}{\|\mathbf b\|^2} \\
&- T \ln\left(\frac{\sum_{t=1}^T|\mathbf b^H \mathbf y_{\mathrm{s},1}(t)|^2}{T\|\mathbf b\|^2}\right).
\end{split}
\end{equation}
Finally, the MLE value of $\theta$ can be obtained by exhaustive searching the feasible region $\left[-\frac{\pi}{2},\frac{\pi}{2}\right]$ to maximize the objective function in \eqref{eq:MLE_theta}. With the above estimator design, it is expected the MSE of this proposed estimator equals to the derived CRB when the sensing SNR is sufficiently large \cite{steven1993fundamentals_estimation,richards2005fundamentals}, which will be verified in the Section~\ref{sec:numerical_results}.

\subsection{CRB Minimization for Sensing-only Scenario}
Next, we explore the minimum achievable CRB for the sensing-only scenario. It is observed from the CRB expression in \eqref{eq:CRB_random_derive} that minimizing $\text{CRB}_1(\theta)$ is equivalent to maximizing $\mathbf a^T\mathbf R_\mathrm{c}\mathbf a^*$. Thus, the CRB minimization problem is formulated as
\begin{subequations}
  \begin{align}\notag
   \text{(P1)}:\max_{\mathbf R_\mathrm{c}} &\quad  \mathbf a^T\mathbf R_\mathrm{c}\mathbf a^*\\ 
    \text { s.t. }&\quad \mathbf R_\mathrm{c} \succeq \mathbf 0,\\ 
    &\quad \mathrm{tr}(\mathbf R_\mathrm{c})\le P, 
  \end{align}
\end{subequations}
for which the optimal solution is given by maximum ratio transmission (MRT) beamformer,\footnote{This optimal beamforming design requires to perfectly know the angle of target w.r.t. the transmitter in advance. This assumption is valid for the target tracking scenario, in which the target's angle can be predicted based on the previous sensing results to facilitate the beamforming design \cite{10138058,9652071}.} i.e., 
\begin{equation}\label{eq:optimal_BF_sensing}
\mathbf R_\mathrm{c}^\star = P\frac{\mathbf a^*\mathbf a^T}{\|\mathbf a\|^2}.
\end{equation}

By substituting the optimal transmit beamforming solution in \eqref{eq:optimal_BF_sensing} into the CRB expression in \eqref{eq:CRB_random_derive}, the minimum CRB for estimating the target's DoA $\theta$  with only Gaussian information signals is
\begin{equation}\label{eq:CRB_min_random}
  \begin{split}
&\text{CRB}_1^\star(\theta)\\
 =& \frac{\sigma_\mathrm{s}^2}{2PT|\alpha|^2\|\mathbf a\|^2\|\dot{\mathbf b}\|^2}\left(1+\frac{\sigma_\mathrm{s}^2}{P|\alpha|^2\|\mathbf b\|^2\|\mathbf a\|^2}\right)\\
=& \frac{3\sigma_\mathrm{s}^2\lambda^2}{2PT\pi^2d_\mathrm{a}^2\cos^2\theta|\alpha|^2M_\mathrm{t}(M_\mathrm{r}-1)M_\mathrm{r}(M_\mathrm{r}+1)}\\
&\times\left(1+\frac{\sigma_\mathrm{s}^2}{P|\alpha|^2M_\mathrm{t}M_\mathrm{r}}\right).
  \end{split}
\end{equation}
Based on the minimum CRB in \eqref{eq:CRB_min_random}, we have the following remark.
\begin{remark}\label{remark:scale_law}
With the optimal transmit beamforming design in \eqref{eq:optimal_BF_sensing}, the resultant CRB decreases inversely proportionally to $P$, $M_\mathrm{t}$, and $M_\mathrm{r}^3$ when the radar sensing SNR is sufficiently large, while it decreases inversely proportionally to $P^2$, $M_\mathrm{t}^2$, and $M_\mathrm{r}^4$ when the radar sensing SNR is small. This scaling behavior is different from the scenario with deterministic signals, in which the minimum CRB, i.e., $\text{CRB}_\mathrm{d}^\star(\theta)=\frac{3\sigma_\mathrm{s}^2\lambda^2}{2PT\pi^2d_\mathrm{a}^2\cos^2\theta|\alpha|^2M_\mathrm{t}(M_\mathrm{r}-1)M_\mathrm{r}(M_\mathrm{r}+1)}$\cite{steven1993fundamentals_estimation,9652071}, decreases inversely proportionally only to $P$, $M_\mathrm{t}$, and $M_\mathrm{r}^3$, respectively. The additional scaling factor that emerges in the Gaussian-signal scenario for low radar sensing SNR explicitly accounts for the performance gap between $\text{CRB}_1^\star(\theta)$ and $\text{CRB}_\mathrm{d}^\star(\theta)$. This gap diminishes explicitly as the sensing SNR increases, eventually converges to zero at sufficiently large sensing SNR.

Notably, increasing the number of receive antennas reduces the CRB more significantly than increasing the number of transmit antennas. This can be intuitively explained as given in the following. First, increasing either type of antenna enhances sensing SNR, lowering the CRB. Besides, since the target's DoA estimation depends on the phase difference across adjacent receive antennas, a larger receive array increases the receive signal dimension, leading directly to improved accuracy in DoA estimation.
\end{remark}

\subsection{SINR-Constrained CRB Minimization}
Finally, we characterize the ISAC performance boundary for the scenario where the BS transmits only Gaussian information signals. In particular, we formulate an optimization problem to minimum the estimation CRB, subject to a minimum SNR requirement at the CU, as well as a maximum transmit power constraint at the BS, which is formulated as
\begin{subequations}
  \begin{align}\notag
   \text{(P2)}:\max_{\mathbf R_\mathrm{c}} &\quad  \mathbf a^T\mathbf R_\mathrm{c}\mathbf a^*\\\label{constr:SINR_random} 
    \text { s.t. }&\quad \frac{\mathbf h^H \mathbf R_\mathrm{c}\mathbf h}{\sigma_\mathrm{c}^2}\ge \gamma_0,\\\label{constr:semi_random}
    & \quad \mathbf R_\mathrm{c} \succeq \mathbf 0,\\\label{constr:power_random} 
    &\quad \mathrm{tr}(\mathbf R_\mathrm{c})\le P.
  \end{align}
\end{subequations}

The above problem (P2.1) is convex and its optimal solution is given in the following proposition.
\begin{proposition}\label{prop:solution_P3}
The optimal solution of (P2) is
\begin{equation}\label{eq:opt_R_c}
\begin{split}
&\mathbf R_\mathrm{c}^\star =\\
 &\begin{cases} 
P\frac{\mathbf a^*\mathbf a^T}{\|\mathbf a\|^2}, \quad &\text{if~} P|\mathbf h^H\mathbf a^*|^2 \ge M_\mathrm{t} \gamma_0\sigma_\mathrm{c}^2,\\
[\mathbf u_1, \mathbf u_2]
\begin{bmatrix}
\lambda_1 & \lambda_{12}\\
\lambda_{12}^* & \lambda_2
\end{bmatrix}
[\mathbf u_1, \mathbf u_2]^H,\!&\text{otherwise},
\end{cases}
\end{split}
\end{equation}
where 
\begin{equation}
\mathbf u_1=\frac{\mathbf h}{\|\mathbf h\|}, \quad \mathbf u_2 = \frac{\mathbf a^*-\mathbf u_1^H\mathbf a^*\mathbf u_1}{\|\mathbf a^*-\mathbf u_1^H\mathbf a^*\mathbf u_1\|},
\end{equation}
\begin{equation}
\lambda_1= \frac{\gamma_0\sigma_\mathrm{c}^2}{\|\mathbf h\|^2}, \quad \lambda_2 = P- \frac{\gamma_0\sigma_\mathrm{c}^2}{\|\mathbf h\|^2}, \quad \lambda_{12} = \sqrt{\lambda_1\lambda_2}\frac{\mathbf u_1^H\mathbf a^*}{|\mathbf u_1^H\mathbf a^*|}.
\end{equation}
\begin{IEEEproof}
Please refer to Appendix~\ref{sub:proof_of_prop_solution_P3}.
\end{IEEEproof}
\end{proposition}

It can be observed from Proposition~\ref{prop:solution_P3} that the optimal solution to Problem (P2), i.e., $\mathbf R_\mathrm{c}^\star$, is rank-one and lies within the subspace spanned by vectors $\mathbf h$ and $\mathbf a^*$. With the optimal transmit beamforming design in \eqref{eq:opt_R_c}, the minimum estimation CRB with a given SNR threshold $\gamma_0$ satisfying $P|\mathbf h^H\mathbf a^*|^2 \ge M_\mathrm{t} \gamma_0\sigma_\mathrm{c}^2$ is
\begin{equation}\label{eq:CRB_rate_random_1}
  \begin{split}
\text{CRB}_1^\mathrm{opt,1}(\theta)
 =&\frac{3\sigma^2\lambda^2}{2PT\pi^2d_\mathrm{a}^2\cos^2\theta|\alpha|^2M_\mathrm{t}(M_\mathrm{r}-1)M_\mathrm{r}(M_\mathrm{r}+1)}\\
&\times\left(1+\frac{\sigma^2}{P|\alpha|^2M_\mathrm{t}M_\mathrm{r}}\right).
  \end{split}
\end{equation}
On the other hand, for the scenario $P|\mathbf h^H\mathbf a^*|^2 < M_\mathrm{t} \gamma_0\sigma_\mathrm{c}^2$, the minimum estimation CRB becomes 
\begin{equation}\label{eq:CRB_rate_random_2}
  \begin{split}
\text{CRB}_1^\mathrm{opt,2}(\theta)
 =&\frac{3\sigma^2\lambda^2}{2T\pi^2d_\mathrm{a}^2\cos^2\theta|\alpha|^2(M_\mathrm{r}-1)M_\mathrm{r}(M_\mathrm{r}+1)E(\gamma_0)}\\
&\times\left(1+\frac{\sigma^2}{|\alpha|^2M_\mathrm{r}E(\gamma_0) }\right),
  \end{split}
\end{equation}
where
\begin{equation}
\begin{split}
E(\gamma_0) = &\left(\frac{\sqrt{\gamma_0\sigma_\mathrm{c}^2}|\mathbf h^H\mathbf a^*|}{\|\mathbf h\|^2}\right.\\
&\left.+\sqrt{\left(P-\frac{\gamma_0\sigma_\mathrm{c}^2}{\|\mathbf h\|^2}\right)^2\left(M_t-\frac{|\mathbf h^H\mathbf a^*|^2}{\|\mathbf h\|^2}\right)^2}\right)^2.
\end{split}
\end{equation}

\begin{remark}\label{remark:scale_law_CRB_rate}
It can be observed from \eqref{eq:CRB_rate_random_1} and \eqref{eq:CRB_rate_random_2} that if $\gamma_0\le \frac{P|\mathbf h^H\mathbf a^*|^2}{M_\mathrm{t}\sigma_\mathrm{c}^2}$, the minimum estimation CRB is constant. Otherwise, for the scenario $\gamma_0> \frac{P|\mathbf h^H\mathbf a^*|^2}{M_\mathrm{t}\sigma_\mathrm{c}^2}$, the term $E(\gamma_0)$ decreases monotonically with the increase in $\gamma_0$ and decays as the square root of the gap $\left(\frac{P\|\mathbf h\|^2}{\sigma_\mathrm{c}^2}-\gamma_0\right)$ near the feasibility boundary $\gamma_0=\frac{P\|\mathbf h\|^2}{\sigma_\mathrm{c}^2}$.
\end{remark}

\section{ISAC with both Gaussian Information and Deterministic Sensing Signals}\label{sec:both_gaussian_and_deterministic}
In this section, we investigate the CRB-rate tradeoff region for the scenario where the BS transmits a superposition of Gaussian information and deterministic sensing signals. In particular, we first derive a closed-form CRB expression for estimating the target DoA. Then, we propose an efficient estimator design for achieving  this bound. Finally, we introduce the transmit beamforming design for both sensing-only and ISAC scenarios.

\subsection{CRB Derivation}
First, we derive the CRB for estimating the target's DoA $\theta$ when the BS transmits both Gaussian information and deterministic sensing signals. In this scenario, we aim to jointly estimate both the target's DoA $\theta$ and the complex channel coefficient $\alpha$. Let $\bm \eta=[\theta, \tilde{\bm\alpha}^T]^T$ denote the parameters that need to be estimated, where $\tilde{\bm\alpha} =[\mathrm {Re}\{\alpha\}, \mathrm {Im}\{\alpha\}]^T$, which includes the target's DoA $\theta$ and the real and imaginary parts of the complex channel coefficient $\alpha$. By stacking the transmitted deterministic sensing signal as $ \mathbf X_0 = \left[\mathbf x_0(1),\cdots,\mathbf x_0(T)\right]$, the received signal over the whole sensing time is modeled as
\begin{equation}\label{eq:Y_2}
\mathbf Y_2 = \alpha \mathbf b \mathbf a^T \mathbf S + \alpha \mathbf b \mathbf a^T \mathbf X_0 +  \mathbf N_\mathrm{s}.
\end{equation}
By further vectorizing \eqref{eq:Y_2} and letting $\tilde{\mathbf x}_0= \mathrm{vec}(\mathbf X_0)$, we have
\begin{equation}\label{eq:echo_y_joint}
\begin{split}
\tilde{\mathbf y}_2&= \mathrm{vec}(\mathbf Y_2)\\
 &= \mathrm{vec}(\alpha \mathbf b \mathbf a^T \mathbf S) + \mathrm{vec}(\alpha \mathbf b \mathbf a^T \mathbf X_0) +  \mathrm{vec}(\mathbf N_\mathrm{s})\\
& = \left(\mathbf I_T \otimes \alpha \mathbf b \mathbf a^T\right)\tilde{\mathbf s} + \left(\mathbf I_T \otimes \alpha \mathbf b \mathbf a^T\right)\tilde{\mathbf x}_0 + \tilde{\mathbf n}_\mathrm{s}\\
& = \tilde{\mathbf u}_\mathrm{c}+ \tilde{\mathbf u}_\mathrm{s} + \tilde{\mathbf n}_\mathrm{s},
\end{split}
\end{equation}
where $\tilde{\mathbf u}_\mathrm{s} = \left(\mathbf I_T \otimes \alpha \mathbf b(\theta) \mathbf a^T(\theta)\right)\tilde{\mathbf x}_0$. For the target sensing receiver, $\tilde{\mathbf x}_0$ is the deterministic signal with predetermined sequence, while $\tilde{\mathbf s}$ is the random signal without specific realizations.

Let $\hat{\mathbf F} \in \mathbb R^{3\times 3}$ denote the FIM  for estimating $\bm \eta$, each element of which is given by \cite{steven1993fundamentals_estimation}
\begin{equation}\label{eq:FIM_both}
\begin{split}
\left[\hat{\mathbf F}\right]_{(i,j)}&=\mathrm{tr}\left(\mathbf R_\mathrm{n}^{-1}\frac{\partial \mathbf R_\mathrm{n}}{\partial \bm \eta_i}\mathbf R_\mathrm{n}^{-1}\frac{\partial \mathbf R_\mathrm{n}}{\partial \bm \eta_j}\right)\\
&\quad +2\mathrm{Re}\left\{\frac{\partial \tilde{\mathbf u}_\mathrm{s}^H}{\partial \bm \eta_i}\mathbf R_\mathrm{n}^{-1}\frac{\partial \tilde{\mathbf u}_\mathrm{s}}{\partial \bm \eta_j}\right\}, \quad i,j\in\{1,2,3\}.
\end{split}
\end{equation} 
Based on \eqref{eq:FIM_both}, we have the following lemma.
\begin{lemma}
The FIM $\hat{\mathbf F}$ for estimating $\bm \eta$  is 
\begin{equation}
\hat{\mathbf F} \!=\! \begin{bmatrix}
\hat F_{\left(\theta,\theta\right)} & \hat{\mathbf F}_{\left(\theta,\tilde{\bm\alpha}\right)} \\
\hat{\mathbf F}_{\left(\theta,\tilde{\bm\alpha}\right)}^T & \hat{\mathbf F}_{\left(\tilde{\bm\alpha},\tilde{\bm\alpha}\right)} 
\end{bmatrix},
\end{equation}
where
\begin{subequations}\label{eq:F_hat_1}
	\begin{align}\notag
\hat{F}_{\left(\theta,\theta\right)}&=\frac{2T|\alpha|^4\left(\mathbf a^T \mathbf R_\mathrm{c} \mathbf a^*\right)^2\|\dot{\mathbf b}\|^2\|\mathbf b\|^2}{\sigma_\mathrm{s}^4(1+\gamma_\text{ran}(\mathbf R_\mathrm{c}))}\\
&\quad+\frac{2T|\alpha |^2\mathbf a^T\mathbf R_\mathrm{s}\mathbf a^*\|\dot{\mathbf b}\|^2}{\sigma_\mathrm{s}^2},\\
\label{eq:F_hat_2}
	\hat{\mathbf F}_{\left(\theta,\tilde{\bm\alpha}\right)}&=\mathbf 0,\\  \notag
\label{eq:F_hat_4}
\hat{\mathbf F}_{\left(\tilde{\bm\alpha},\tilde{\bm\alpha}\right)}&=\frac{4T\left(\mathbf a^T \mathbf R_\mathrm{c} \mathbf a^*\right)^2 \|\mathbf b\|^4}{\sigma_\mathrm{s}^4(1+\gamma_\text{ran}(\mathbf R_\mathrm{c}))^2}
\begin{bmatrix}
\mathrm{Re}\{\alpha\} \\
\mathrm{Im}\{\alpha\} 
\end{bmatrix}
\begin{bmatrix}
\mathrm{Re}\{\alpha\} &
\mathrm{Im}\{\alpha\} 
\end{bmatrix}\\
&\quad+\frac{2T\mathbf a^T\mathbf R_\mathrm{s}\mathbf a^*\|\mathbf b\|^2}{\sigma_\mathrm{s}^2(1+\gamma_\text{ran}(\mathbf R_\mathrm{c}))}\mathbf I_2.
		\end{align}
\end{subequations}
\begin{IEEEproof}
Please refer to Appendix~\ref{sec:the_derivation_of_the_fim_mathbf_hat_f_}.
\end{IEEEproof}
\end{lemma}

Finally, the CRB for estimating the target's DoA $\theta$ when the BS transmits a superposition of Gaussian information and deterministic sensing signals is given by the first diagonal element of $\hat{\mathbf F}^{-1}$, i.e.,
\begin{equation}\label{eq:CRB_joint}
\begin{split}
\text{CRB}_2(\theta) =& \left[\hat{\mathbf F}^{-1}\right]_{(1,1)}
= \left(\hat F_{\left(\theta,\theta\right)} - \hat{\mathbf F}_{\left(\theta,\tilde{\bm\alpha}\right)}
\hat{\mathbf F}_{\left(\tilde{\bm\alpha},\tilde{\bm\alpha}\right)}^{-1}
\hat{\mathbf F}_{\left(\theta,\tilde{\bm\alpha}\right)}^T\right)^{-1}\\
=& \frac{\sigma_\mathrm{s}^2}{2T|\alpha|^2f(\mathbf R_\mathrm{c}, \mathbf R_\mathrm{s})},
\end{split}
\end{equation}
where 
\begin{equation}\label{eq:CRB_joint_denominator}
\begin{split}
 f(\mathbf R_\mathrm{c}, \mathbf R_\mathrm{s})
 &= \mathbf a^T\mathbf R_\mathrm{s}\mathbf a^*\|\dot{\mathbf b}\|^2\\
 &\quad + \frac{|\alpha|^2\mathbf a^T\mathbf R_\mathrm{c}\mathbf a^*\|\mathbf b\|^2/\sigma_\mathrm{s}^2}{1+|\alpha|^2\mathbf a^T\mathbf R_\mathrm{c}\mathbf a^*\|\mathbf b\|^2/\sigma_\mathrm{s}^2}\mathbf a^T \mathbf R_\mathrm{c} \mathbf a^*\|\dot{\mathbf b}\|^2.
\end{split}
\end{equation}

\begin{remark}
By comparing the CRBs in \eqref{eq:CRB_random_derive}, \eqref{eq:CRB_deter_derive}, and \eqref{eq:CRB_joint}, the same beamforming design at the BS results
\begin{equation}
\text{CRB}_\mathrm{d}(\theta) \le \text{CRB}_2(\theta)\le \text{CRB}_1(\theta).
\end{equation}
The explanation is given in the following. We assume the transmit beamforming design is identical across scenarios with Gaussian-only signals, deterministic-only signals, or a superposition of both, i.e., $\mathbf R_\mathrm{c} =\mathbf R_\mathrm{x}$, $\mathbf R_\mathrm{s}=\mathbf R_\mathrm{x}$, and $\mathbf R'_\mathrm{c}+\mathbf R'_\mathrm{s}=\mathbf R_\mathrm{x}$ for these three scenarios, respectively. In this case, we have
\begin{equation}
\begin{split}
&\frac{|\alpha|^2\mathbf a^T\mathbf R_\mathrm{c}\mathbf a^*\|\mathbf b\|^2/\sigma_\mathrm{s}^2}{1+|\alpha|^2\mathbf a^T\mathbf R_\mathrm{c}\mathbf a^*\|\mathbf b\|^2/\sigma_\mathrm{s}^2}\mathbf a^T \mathbf R_\mathrm{c} \mathbf a^*\|\dot{\mathbf b}\|^2\\
\le~ & \mathbf a^T\mathbf R'_\mathrm{s}\mathbf a^*\|\dot{\mathbf b}\|^2 \\
&+ \frac{|\alpha|^2\mathbf a^T\mathbf R'_\mathrm{c}\mathbf a^*\|\mathbf b\|^2/\sigma_\mathrm{s}^2}{1+|\alpha|^2\mathbf a^T\mathbf R'_\mathrm{c}\mathbf a^*\|\mathbf b\|^2/\sigma_\mathrm{s}^2}\mathbf a^T \mathbf R'_\mathrm{c} \mathbf a^*\|\dot{\mathbf b}\|^2\\
\le~&\mathbf a^T\mathbf R_\mathrm{s}\mathbf a^*\|\dot{\mathbf b}\|^2.
\end{split}
\end{equation}
Thus, the CRB with a superposition of Gaussian and deterministic signals is smaller than that with Gaussian-only signals, and larger than that with deterministic-only signals.
\end{remark}

\subsection{Practical Estimator Design}
In this subsection, we design a practical MLE-based estimator when the BS transmits a superposition of Gaussian information and deterministic sensing signals. Based on the received echo signal model in \eqref{eq:echo_y_joint}, the likelihood function of the received signal is 
\begin{equation}\label{eq:PDF_joint}
p_2(\tilde{\mathbf y}_2) = \frac{\exp\left(-(\tilde{\mathbf y}_2-\tilde{\mathbf u}_\mathrm{s})^H\mathbf R_\mathrm{n}^{-1} (\tilde{\mathbf y}_2-\tilde{\mathbf u}_\mathrm{s})\right)}{\pi^{M_\mathrm{t}T}\det(\mathbf R_\mathrm{n})}.
\end{equation}
In this scenario, we aim to estimate both the target's DoA $\theta$ and the complex channel coefficient $\alpha$. Due to the complex expression of the likelihood function in \eqref{eq:PDF_joint}, it is mathematically challenging to obtain a closed-form MLE for $\alpha$ under a given $\theta$ by directly solving the partial derivate of $\ln p_2(\tilde{\mathbf y}_2)$ w.r.t. $\alpha$ as done in Section~\ref{sub:practical_estimator_design_1}. As an alternative, we propose a heuristic estimator that first estimate the phase and magnitude of $\alpha$, respectively, and then construct the estimation value of $\alpha$ relying on the estimated phase and magnitude. In particular, we explore the deterministic sensing signal to estimate the phase of the complex channel coefficient $\alpha$, by matching the received signals with the transmitted deterministic signals with known realizations \cite{10138058,TANG20131349}, which is given as
\begin{equation}\label{eq:estimator_alpha_angle}
\phi_{\alpha_\text{MLE}} = \arg\left(\sum_{t=1}^T\mathbf s^H(t)\mathbf a^*\mathbf b^H\mathbf y_{\mathrm{s},2}(t)  \right).
\end{equation}
Furthermore, we utilize both random and deterministic signals to estimate the magnitude of $\alpha$ as that in Section~\ref{sub:practical_estimator_design_1}, i.e.,
\begin{equation}\label{eq:estimator_alpha_norm}
|\alpha|_\text{MLE} = \sqrt{\frac{\frac{1}{T}\sum_{t=1}^T|\mathbf b^H\mathbf y_{\mathrm{s},2}(t)|^2}{\|\mathbf b\|^4\mathbf a^T(\mathbf R_\mathrm{c}+\mathbf R_\mathrm{s})\mathbf a^*}-\frac{\sigma^2}{\|\mathbf b\|^2\mathbf a^T(\mathbf R_\mathrm{c}+\mathbf R_\mathrm{s})\mathbf a^*}}.
\end{equation}
Combining the estimators in \eqref{eq:estimator_alpha_angle} and $\eqref{eq:estimator_alpha_norm}$, the channel coefficient $\alpha$ under any given DoA $\theta$ and DoD $\phi$ can be estimated as
\begin{equation}\label{eq:alpha_joint}
\alpha_\text{MLE}(\theta,\phi) = |\alpha_\text{MLE}| \exp(\jmath\phi_{\alpha_\text{MLE}}).
\end{equation}
Finally, by substituting the closed-form estimated value of $\alpha$ from \eqref{eq:alpha_joint} into \eqref{eq:PDF_joint}, the target's DoA $\theta$ can be effectively estimated by exhaustive searching over the feasible region $\left[-\frac{\pi}{2},\frac{\pi}{2}\right]$ to maximize the likelihood function value\cite{TANG20131349}. Simulation results in Section~\ref{sec:numerical_results} will show that the MSE of the proposed estimator approaches the derived CRB when the sensing SNR is sufficiently large.

\subsection{CRB Minimization for Sensing-only Scenario}

Next, we explore the optimal CRB minimization for the sensing-only scenario, where the BS transmits a superposition of Gaussian information and deterministic sensing signals. Based on the CRB expression in \eqref{eq:CRB_joint}, the CRB minimization problem is formulated as 
\begin{subequations}
  \begin{align}\notag
   \text{(P3)}:\max_{\mathbf R_\mathrm{c}, \mathbf R_\mathrm{s}} &\quad  \mathbf a^T\mathbf R_\mathrm{s}\mathbf a^*\|\dot{\mathbf b}\|^2\\\notag
 &\quad + \frac{|\alpha|^2\mathbf a^T\mathbf R_\mathrm{c}\mathbf a^*\|\mathbf b\|^2/\sigma_\mathrm{s}^2}{1+|\alpha|^2\mathbf a^T\mathbf R_\mathrm{c}\mathbf a^*\|\mathbf b\|^2/\sigma_\mathrm{s}^2}\mathbf a^T \mathbf R_\mathrm{c} \mathbf a^*\|\dot{\mathbf b}\|^2\\ \label{constr:semi_sensing}
    \text { s.t. }&\quad \mathbf R_\mathrm{c} \succeq \mathbf 0, \mathbf R_\mathrm{s} \succeq \mathbf 0,\\ \label{constr:power_sensing}  
    &\quad \mathrm{tr}(\mathbf R_\mathrm{c})+ \mathrm{tr}(\mathbf R_\mathrm{s})\le P.
  \end{align}
\end{subequations}
Next, we have the following proposition.
\begin{proposition} \label{pro:R_c}
The optimality of (P3) is attained when $\mathbf R_\mathrm{c}^\star =\mathbf 0$, i.e., no power is allocated to the Gaussian information signal for sensing-only optimization.
\end{proposition}
\begin{IEEEproof}
First, we assume that $\hat{\mathbf R}_\mathrm{c}$ and $\hat{\mathbf R}_0$ are the optimal solution of problem (P3) with $\hat{\mathbf R}_\mathrm{c} \succ \mathbf 0$. Then,
 we can always reconstruct an alternative solution of problem (P3) as $\mathbf R_\mathrm{c}^\star=\mathbf 0$ and $\mathbf R_\mathrm{s}^\star=\hat{\mathbf R}_0 + \hat{\mathbf R}_\mathrm{c}$. It is clear that the reconstructed solution satisfies constraints \eqref{constr:semi_sensing}~\text{and}~\eqref{constr:power_sensing}, and achieves a higher objective value.
This contradicts the assumption that $\hat{\mathbf R}_\mathrm{c}$ and $\hat{\mathbf R}_0$ are the optimal solution of problem (P3). Thus, the optimal solution of problem (P3) should satisfy that $\mathbf R_\mathrm{c}^\star =\mathbf 0$.
\end{IEEEproof}

By applying Proposition~\ref{pro:R_c}, the CRB minimization problem (P3) becomes
\begin{subequations}
  \begin{align}\notag
   \text{(P3.1)}:\max_{\mathbf R_\mathrm{s}} &\quad \mathbf a^T\mathbf R_\mathrm{s}\mathbf a^*\|\dot{\mathbf b}\|^2\\ 
    \text { s.t. }    &\quad \mathbf R_\mathrm{s} \succeq \mathbf 0,\\ 
    &\quad \mathrm{tr}(\mathbf R_\mathrm{s})\le P.  
  \end{align}
\end{subequations}
Obviously, the optimality of problem (P3.1) is achieved with MRT beamformer design, i.e., $\mathbf R_\mathrm{s}^\star = P\frac{\mathbf a^*\mathbf a^T}{\|\mathbf a\|^2}$. Thus, by utilizing Proposition~\ref{pro:R_c}, the optimal solution to problems (P3) is given in the following proposition.
\begin{proposition}\label{prop:(P3))}
The optimal solution to problem (P3) is 
\begin{subequations}
\begin{align}\mathbf R_\mathrm{s}^\star &= P\frac{\mathbf a^*\mathbf a^T}{\|\mathbf a\|^2},\\
 \mathbf R_\mathrm{c}^\star&=\mathbf 0.
\end{align}
\end{subequations}
\end{proposition}

With the optimal beamforming design in Proposition~\ref{prop:(P3))}, the resultant minimum CRB is
\begin{equation}\label{eq:CRB_min_joint}
  \begin{split}
&\text{CRB}_2^\star(\theta)\\
=& \frac{\sigma^2}{2PT|\alpha|^2\|\mathbf a\|^2\|\dot{\mathbf b}\|^2}\\
=& \frac{3\sigma^2\lambda^2}{2PT\pi^2d_\mathrm{a}^2\sin^2\theta|\alpha|^2M_\mathrm{t}(M_\mathrm{r}-1)M_\mathrm{r}(M_\mathrm{r}+1)}.
  \end{split}
\end{equation}
Based on Proposition~\ref{prop:(P3))} and the minimum CRB in \eqref{eq:CRB_min_joint}, we have the following remark.
\begin{remark}
When the BS transmits a superposition of Gaussian information and deterministic sensing signals, the minimum estimation CRB is achievable by exclusively transmitting deterministic sensing signals. This result confirms our understanding that deterministic signals are preferable to the random signal for sensing tasks, aligning the previous literature \cite{10147248,10206462,10471902,10596930,10645253,xie2024sensing}. Meanwhile, it is also observed that the minimum CRB in \eqref{eq:CRB_min_joint} decreases inversely proportionally to $P$, $M_\mathrm{t}$, and $M_\mathrm{r}^3$, respectively.
\end{remark}

\subsection{SINR-Constrained CRB Minimization}

Next, we explore the CRB-SINR performance region when the BS transmits a superposition of Gaussian information and deterministic sensing signals. Based on the CRB expression in \eqref{eq:CRB_joint}, the SINR-constrained CRB minimization problem is formulated as
\begin{subequations}
  \begin{align}\notag
   \text{(P4)}:\max_{\mathbf R_\mathrm{c}, \mathbf R_\mathrm{s}} &\quad  \mathbf a^T\mathbf R_\mathrm{s}\mathbf a^*\|\dot{\mathbf b}\|^2\\\notag
 &\quad + \frac{|\alpha|^2\mathbf a^T\mathbf R_\mathrm{c}\mathbf a^*\|\mathbf b\|^2/\sigma_\mathrm{s}^2}{1+|\alpha|^2\mathbf a^T\mathbf R_\mathrm{c}\mathbf a^*\|\mathbf b\|^2/\sigma_\mathrm{s}^2}\mathbf a^T \mathbf R_\mathrm{c} \mathbf a^*\|\dot{\mathbf b}\|^2\\\label{constr:SINR}
    \text { s.t. } &\quad \frac{\mathbf h^H \mathbf R_\mathrm{c}\mathbf h}{\mathbf h^H\mathbf R_\mathrm{s}\mathbf h + \sigma_\mathrm{c}^2}\ge \gamma_0,\\\label{constr:semi}
    &\quad \mathbf R_\mathrm{c} \succeq \mathbf 0, \mathbf R_\mathrm{s} \succeq \mathbf 0,\\ \label{constr:power}  
    &\quad \mathrm{tr}(\mathbf R_\mathrm{c})+ \mathrm{tr}(\mathbf R_\mathrm{s})\le P.
  \end{align}
\end{subequations}
The objective function of (P4) is equivalently expressed as
\begin{equation}\label{eq:f_CRB}
\begin{split}
  f(\mathbf R_\mathrm{c}, \mathbf R_\mathrm{s})
 &=\mathbf a^T\mathbf R_\mathrm{s}\mathbf a^*\|\dot{\mathbf b}\|^2 + \mathbf a^T\mathbf R_c\mathbf a^*\|\dot{\mathbf b}\|^2
-\frac{\sigma_\mathrm{s}^2\|\dot{\mathbf b}\|^2}{|\alpha|^2\|\mathbf b\|^2}\\
&\quad+\frac{\sigma_\mathrm{s}^2\|\dot{\mathbf b}\|^2}{|\alpha|^2\|\mathbf b\|^2}\frac{1}{1+|\alpha|^2\mathbf a^T\mathbf R_\mathrm{c}\mathbf a^*\|\mathbf b\|^2/\sigma_\mathrm{s}^2}\\
&= f_1(\mathbf R_\mathrm{c},\mathbf R_\mathrm{s}) + f_2(\mathbf R_\mathrm{c}),
\end{split}
\end{equation}
where $f_1(\mathbf R_\mathrm{c},\mathbf R_\mathrm{s}) = \mathbf a^T\mathbf R_\mathrm{s}\mathbf a^*\|\dot{\mathbf b}\|^2 + \mathbf a^T\mathbf R_c\mathbf a^*\|\dot{\mathbf b}\|^2-\frac{\sigma_\mathrm{s}^2\|\dot{\mathbf b}\|^2}{|\alpha|^2\|\mathbf b\|^2}$ and $f_2(\mathbf R_\mathrm{c})=\frac{\sigma_\mathrm{s}^2\|\dot{\mathbf b}\|^2}{|\alpha|^2\|\mathbf b\|^2}\frac{1}{1+|\alpha|^2\mathbf a^T\mathbf R_\mathrm{c}\mathbf a^*\|\mathbf b\|^2/\sigma_\mathrm{s}^2}$. In \eqref{eq:f_CRB}, $f_1(\mathbf R_\mathrm{c},\mathbf R_\mathrm{s})$ is affine and $f_2(\mathbf R_\mathrm{c})$ is convex.

To efficiently address problem (P4), we introduce an SCA algorithm \cite{9916163} by first substituting the convex component $f_2(\mathbf R_\mathrm{c})$ with its lower bound surrogate function leveraging a series of first-order Taylor expansions, and then solving the reformulated problems iteratively. In particular, let $\mathbf R_\mathrm{c}^{(k)}$ denote the local point of $\mathbf R_\mathrm{c}$ in iteration $k$ of SCA. The first-order Taylor expansion of $f_2(\mathbf R_\mathrm{c})$ in  iteration $k$ is given as
\begin{equation}
\begin{split}
f_2(\mathbf R_\mathrm{c}) &\ge \frac{\sigma_\mathrm{s}^2\|\dot{\mathbf b}\|^2}{|\alpha|^2\|\mathbf b\|^2}\frac{1}{1+|\alpha|^2\mathbf a^T\mathbf R_\mathrm{c}^{(k)}\mathbf a^*\|\mathbf b\|^2/\sigma_\mathrm{s}^2}\\
&\quad -\|\dot{\mathbf b}\|^2\frac{\mathbf a^T \left(\mathbf R_\mathrm{c}-\mathbf R_\mathrm{c}^{(k)}\right)\mathbf a^* }{\left(1\!+\!|\alpha|^2\mathbf a^T\mathbf R_\mathrm{c}^{(k)}\mathbf a^*\|\mathbf b\|^2/\sigma_\mathrm{s}^2\right)^2}\\
&\triangleq \tilde{f}_2^{(k)}(\mathbf R_\mathrm{c}).
\end{split}
\end{equation}
Obviously, function $\tilde{f}_2(\mathbf R_\mathrm{c})$ is affine w.r.t. $\mathbf R_\mathrm{c}$.
By substituting the convex component $f_2(\mathbf R_\mathrm{c})$ with its first-order Taylor expansion, a suboptimal solution to problem (P4) can be obtained by solving as a series of problems (P4,$k$), i.e.,
\begin{subequations}
  \begin{align}\notag
   \text{(P4,$k$)}:\max_{\mathbf R_\mathrm{c}, \mathbf R_\mathrm{s}} &\quad  f_1(\mathbf R_\mathrm{c}, \mathbf R_\mathrm{s})+\tilde{f}_2^{(k)}(\mathbf R_\mathrm{c})\\\notag
    \text { s.t. } &\quad \eqref{constr:SINR},~\eqref{constr:semi},~\text{and}~\eqref{constr:power}.
  \end{align}
\end{subequations}
Note that problem (P4,$k$) is convex and thus can be optimally solved by standard convex programing solvers such as CVX\cite{cvx}. After obtaining the optimal solution to problem (P4,$k$), which is denoted by $\mathbf R_\mathrm{c}^\mathrm{opt}$ and $\mathbf R_\mathrm{s}^\mathrm{opt}$, the local point of $\mathbf R_\mathrm{c}$ in iteration $k+1$ is updated by $\mathbf R_\mathrm{c}^{(k+1)}=\mathbf R_\mathrm{c}^\mathrm{opt}$. By iteratively solving the series of convex problems (P4,$k$), an local optimal solution of problem (P4) is obtained. The convergence of the SCA algorithm is ensured because $\tilde{f}_2^{(k)}(\mathbf R_\mathrm{c})$ is a lower bound of $f_2(\mathbf R_\mathrm{c})$. In particular, the objective function of problem (P4) with the solution in iteration $k+1$ is larger than that in iteration $k$, i.e.,
\begin{equation}
\begin{split}
&\quad f_1(\mathbf R_\mathrm{c}^{(k+1)}, \mathbf R_\mathrm{s}^{(k+1)})+ f_2(\mathbf R_\mathrm{c}^{(k+1)})\\
&\ge f_1(\mathbf R_\mathrm{c}^{(k+1)}, \mathbf R_\mathrm{s}^{(k+1)})+\tilde{f}_2^{(k)}(\mathbf R_\mathrm{c}^{(k+1)})\\
&\ge f_1(\mathbf R_\mathrm{c}^{(k)}, \mathbf R_\mathrm{s}^{(k)})+\tilde{f}_2^{(k)}(\mathbf R_\mathrm{c}^{(k)})\\
&= f_1(\mathbf R_\mathrm{c}^{(k)}, \mathbf R_\mathrm{s}^{(k)})+ f_2(\mathbf R_\mathrm{c}^{(k)}).
\end{split}
\end{equation}
In summary, the detailed algorithm for problem (P4) is given in Algorithm~\ref{tab:table1}.

\begin{algorithm}[t]
\caption{Proposed Algorithm for Solving Problem (P4)}
\label{tab:table1}
\begin{algorithmic}[1] 
\STATE Set iteration index $k= 1$ and initialize  $\mathbf R_\mathrm{c}^{(k)}$.
  \REPEAT
  \STATE Construct $\tilde{f}_2^{(k)}(\mathbf R_\mathrm{c})$ given $\mathbf R_\mathrm{c}^{(k)}$ and solve problem (P4,$k$) to obtain $\mathbf R_\mathrm{c}^\mathrm{opt}$ and $\mathbf R_\mathrm{s}^\mathrm{opt}$.\\
  \STATE Update $\mathbf R_\mathrm{c}^{(k+1)}\gets\mathbf R_\mathrm{c}^\mathrm{opt}$.
   \STATE Update $k\gets k+1$.
  \UNTIL the convergence criterion is met or the maximum number of outer iterations is reached.
\end{algorithmic}
\end{algorithm}

\section{Numerical Results}\label{sec:numerical_results} 
In this section, we present numerical results to validate the correctness of CRB derivations and demonstrate the performance of the proposed transmit beamforming design. The channel between the BS and the CU is modeled as a Rician fading channel, i.e., 
\begin{equation}
\mathbf h = \sqrt{\frac{K_c}{K_c+1}} \mathbf h_\text{los} + \sqrt{\frac{1}{K_c+1}} \mathbf h_\text{nlos}, 
\end{equation}
where $\mathbf h_\text{los}\in \mathbb C^{M_\mathrm{t}\times  1}$ and $\mathbf h_\text{nlos}\in \mathbb C^{M_\mathrm{t}\times  1}$ denote the line-of-sight (LoS) and Rayleigh fading components, respectively, and $K_c=1$ denotes the Rician factor. The path loss model is given as
\begin{equation}
L(d) = K_0 (d/d_0)^{-\alpha_0},
\end{equation}
where $d$ is the distance of the transmission link, $K_0 =-30$~dB is the path loss at distance $d_0=1$~m, and $\alpha_0=2.5$ is the path-loss exponent\cite{8811733}. The transmission distance of the BS-target, target-receiver, and BS-CU links are set as $d_\text{BT}=200$~m, $d_\text{TR}=200$~m, and $d_\text{BC}=1000$~m, respectively. The target's DoA w.r.t. the sensing receiver is set as $\theta=0$. The numbers of transmit and receive antennas are set as $M_\mathrm{t}=M_\mathrm{r}=32$. The transmit power at the BS is set as $P = 30$~dBm. The power of noise at the CU and sensing receiver are set as $\sigma_\mathrm{c}^2=\sigma_\mathrm{s}^2=-80$~dBm. Meanwhile, we set the target reflection coefficient $\beta$ to unity and the length of symbols as $T=1024$.

\subsection{Sensing-Only Performance Analysis}
First, we present simulation results for the sensing-only scenario. For performance comparison, we consider the following benchmark schemes.
\subsubsection{Target Sensing with Deterministic Signals} In this scheme, the BS transmits deterministic  signals to estimate the target's DoA and the sample covariance matrix of transmit signal is set as $\mathbf R_\mathrm{s} = P\frac{\mathbf a^* \mathbf a^T}{\|\mathbf a\|^2}$ to minimize the estimation CRB in \eqref{eq:CRB_deter_derive}.
\subsubsection{ISAC Signal with Both Gaussian and Deterministic Signals} In this scheme, the covariance matrix of Gaussian signals and the sample covariance matrix of deterministic  signals are set as $\mathbf R_\mathrm{c} = P_1\frac{\mathbf a^* \mathbf a^T}{\|\mathbf a\|^2}$ and $ \mathbf R_\mathrm{s} = P_2\frac{\mathbf a^* \mathbf a^T}{\|\mathbf a\|^2}$, respectively, with $P_1+P_2 =P$, where $P_1$ and $P_2$ are the power allocated to the Gaussian and deterministic signals, respectively.

\begin{figure}[t]
        \centering
        \includegraphics[width=0.45\textwidth]{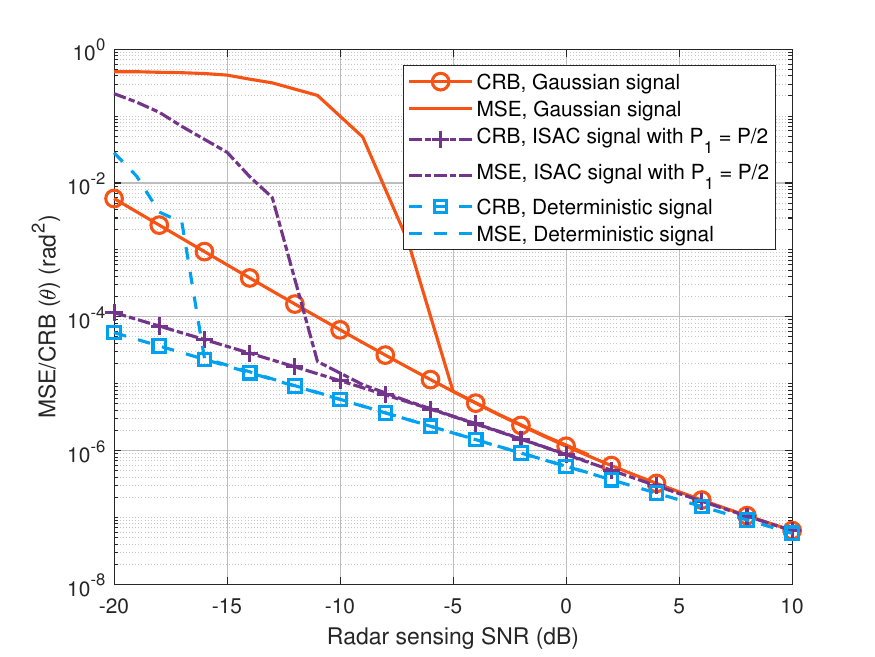}
        \caption{The CRB for estimating the target's DoA versus the radar sensing SNR.}
        \label{fig:CRB_radar_SNR}
\end{figure}

Fig.~\ref{fig:CRB_radar_SNR} shows the CRB performance for estimating the target's DoA versus the radar sensing SNR across different transmit signal models. It is observed that the estimation CRBs consistently decrease as the radar sensing SNR increases. It is also shown that the performance gap between Gaussian-only and deterministic-only decreases with the radar sensing SNR, and the CRB performance with Gaussian information signals approaches that of deterministic signals at sufficiently high sensing SNR. This shows the feasibility of utilizing Gaussian information signals to achieve accurate target sensing in practical ISAC systems when the noise power is low, highlighting the crucial role of deterministic signals in decreasing estimation CRB when the noise power is high. Furthermore, in the high SNR region, the estimation MSEs converge to the corresponding derived CRBs, which validates the correctness of our CRB derivations \cite{steven1993fundamentals_estimation,richards2005fundamentals}. Meanwhile, it is also shown that the MSE performance with ISAC signals is higher than that with deterministic sensing signals and lower than that with Gaussian information signals, which shows the significance of deterministic signals for practical sensing performance enhancement.

\begin{figure}[t]
        \centering
        \includegraphics[width=0.45\textwidth]{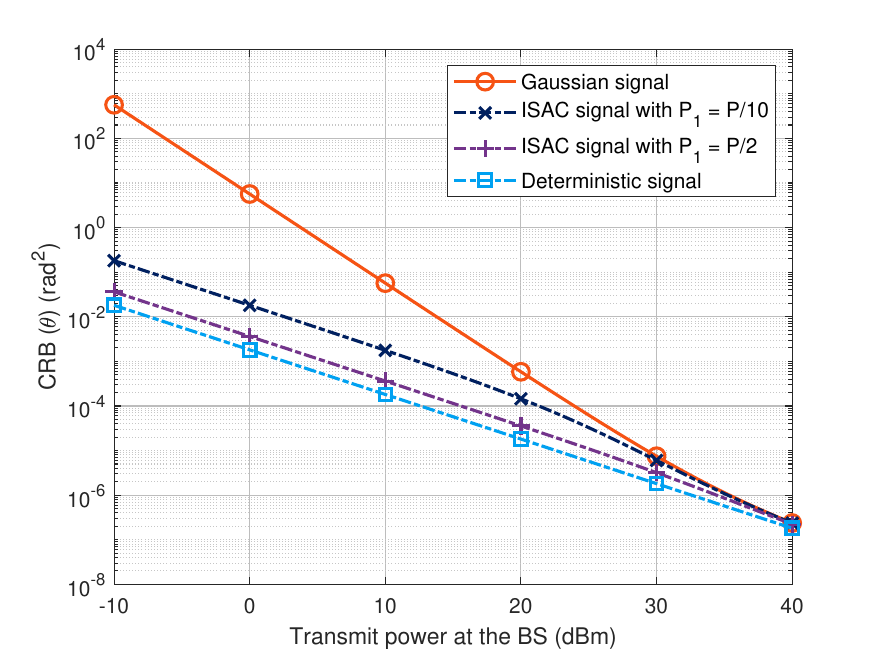}
        \caption{The CRB for estimating the target's DoA versus the transmit power at the BS.}
        \label{fig:CRB_power}
\end{figure}

Fig.~\ref{fig:CRB_power} illustrates the CRB for estimating the target's DoA versus the transmit power $P$ at the BS. For the Gaussian-only scenario, it is shown that increasing the transmit power $P$ from $-10$~dBm to $0$~dBm and from $30$~dBm to $40$~dBm lead to estimation CRB reductions of $19.96$~dB and $10.36$~dB decrease, respectively, indicating diminishing returns at higher power loads. This result is consistent with the scaling law analysis in Remarks~\ref{remark:scale_law}, i.e., the estimation CRB decreases inversely proportionally to $P^2$ and $P$ in the low and high SNR regions, respectively. For the deterministic-only scenario, increasing the transmit power $P$ from $-10$~dBm to $0$~dBm or from $30$~dBm to $40$~dBm all result in $10$~dB CRB reduction.

\begin{figure}[t]
        \centering
        \includegraphics[width=0.45\textwidth]{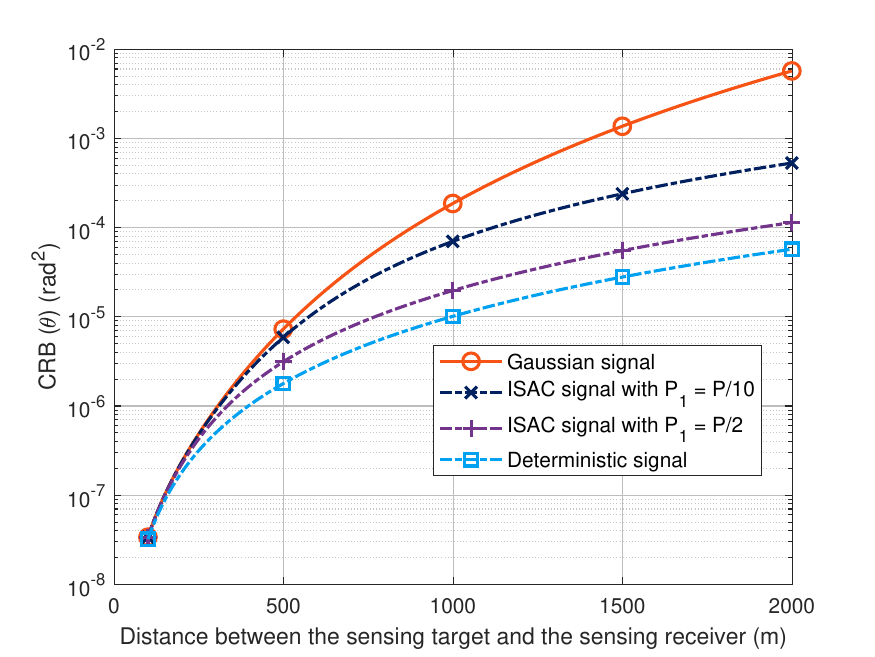}
        \caption{The CRB for estimating the target's DoA versus the distance between the sensing target and the sensing receiver.}
        \label{fig:CRB_distance}
\end{figure}

Fig.~\ref{fig:CRB_distance} shows the CRB for estimating the target's DoA versus the distance between the sensing target and the sensing receiver. It is observed that the estimation CRBs increase with enlarging sensing distance. Under the same estimation CRB requirement, the maximum sensing distance with Gaussian signals is less than that with ISAC or deterministic signals. This is due to the fact that the CRB performance gap between Gaussian-only scenario and deterministic-only scenario becomes more significant when the sensing SNR is small. Meanwhile, the utilization of deterministic signals, even when the power of deterministic signals is as low as $P_1=P/10$, is able to significantly enlarge the maximum sensing distance under the same CRB requirement.

\subsection{CRB-Rate Tradeoff}
Next, we analyze the tradeoff between the estimation CRB and the achievable communication rate. For a comprehensive performance comparison, we consider the following benchmark schemes. 

 \subsubsection{ISAC with Given Realizations of Information Signal\cite{9652071}} In this scheme, we assume that the realizations of information and sensing signals are all perfectly known by the sensing system, and the sensing duration is sufficiently long such that the sample covariance matrix of the information signal can be accurately approximated by its covariance matrix. Thus, the deterministic-signal CRB given in \eqref{eq:CRB_deter_derive} is utilized as the sensing performance metric. The transmit beamforming at the BS is optimized to minimize the estimation CRB subject to an SINR constraint at the CU\cite{9652071}.
\subsubsection{Time Switching} 
In this scheme, the BS alternates between sensing and communication modes. Specifically, the beamforming design in each mode is optimized to either minimize the estimation CRB or maximize the communication rate, respectively, i.e, $\mathbf R_\mathrm{c} = P\mathbf h \mathbf h^H/\|\mathbf h\|^2$ or $\mathbf R_\mathrm{s} = P\mathbf a^* \mathbf a^T/\|\mathbf a\|^2$. Let $T_1$ and $T_2$ denote the time allocated to sensing and communication, respectively, with $T_1+T_2 = T$. The resulting estimation CRB is defined as the CRB with $T_1$ sensing symbols and the resulting communication rate is given as the average rate over all $T$ symbols, i.e., $R' = T_1/T R_1$. Through adjusting the allocated time for sensing and communication, various boundary points can be achieved to balance the sensing and communication performance.
\subsubsection{Power Splitting} Let $P_1$ and $P_2$ denote the power allocated to sensing and communication, respectively, with $P_1+P_2 = P$. The transmit beamformer at the BS is set as $\mathbf R_\mathrm{x} = P_1\mathbf h \mathbf h^H/\|\mathbf h\|^2 +  P_2\mathbf a^* \mathbf a^T/\|\mathbf a\|^2$. The tradeoff between sensing and communication is achievable by adjusting the power allocated for sensing and communication.

\begin{figure}[t]
        \centering
        \includegraphics[width=0.45\textwidth]{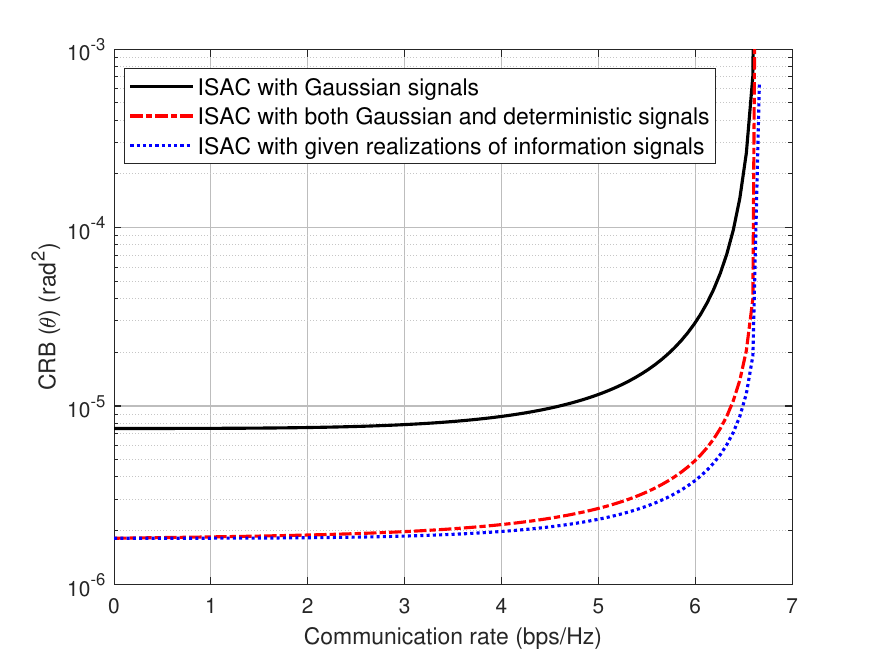}
        \caption{The CRB for estimating the target's DoA versus the communication rate across different transmit signal models.}
        \label{fig:CRB_rate}
\end{figure}

Fig.~\ref{fig:CRB_rate} illustrates the CRB-rate tradeoff across different transmit signal models when exploiting the proposed transmit beamforming design. It is shown that the estimation CRB increases with increasing communication rate requirements, which shows the tradeoff between sensing and communication. Transmitting only Gaussian signals increases the CRB relative to the benchmark with known signal realizations, as the unknown signal sequence introduces uncertainty. However, the ISAC scheme employing both Gaussian and deterministic signals achieves a significant CRB reduction compared to Gaussian-only transmission, yielding performance close to that of the benchmark scheme with given realizations of information signals.

\begin{figure}[t]
        \centering
        \includegraphics[width=0.45\textwidth]{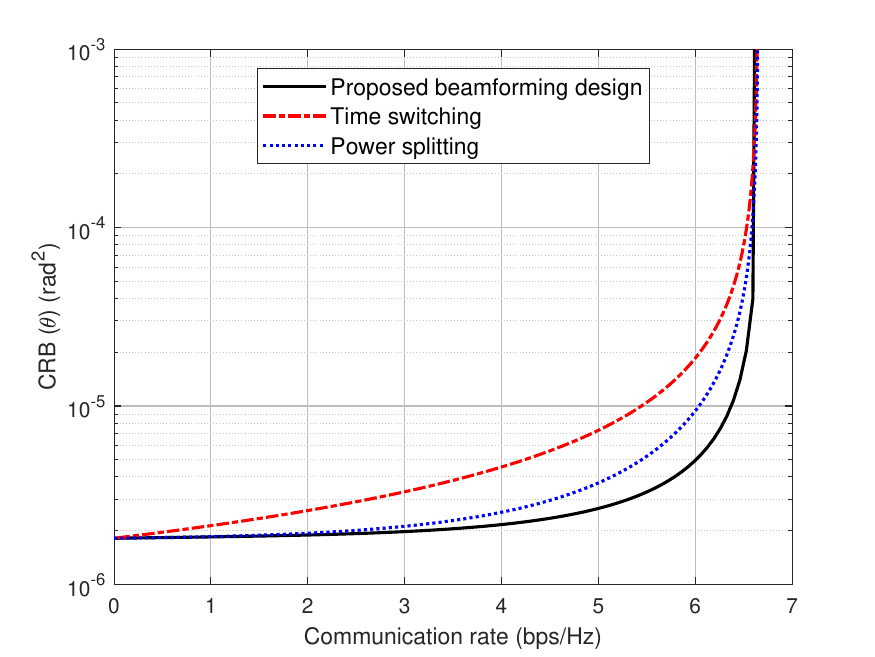}
        \caption{ The CRB for estimating the target's DoA versus the communication rate across different beamforming designs.}
        \label{fig:CRB_BF}
\end{figure}

Fig.~\ref{fig:CRB_BF} shows the CRB-rate performance tradeoff across various beamforming design methods. It can be observed that the proposed design achieves a lower estimation CRB compared to time-switching and power-splitting benchmark schemes and the performance enhancement is notable when the communication rate requirement is moderate. This demonstrates the great benefits of beamforming design for ISAC performance enhancement.

\section{Conclusion}\label{sec:conclusion}
This paper investigated a bistatic ISAC system consisting of a BS, a sensing receiver, a CU, and a point target. The BS delivers information symbols to the CU and the sensing receiver estimates the target's DoA by processing the received target's echo signals reflected from the target. We considered two distinct transmission strategies, in which the BS transmits either only Gaussian information or a superposition of Gaussian information and deterministic sensing signals. Due to the inherent randomness of the Gaussian information signals, the sensing system was assumed to have knowledge of the covariance matrix of Gaussian information signals  instead of their specific realizations. Under these scenarios, we first derived closed-form CRB expressions and introduced two practical estimators. Then, we optimized the transmit beamforming at the BS to minimize the estimation CRBs for the sensing-only scenario and explored the estimation CRB-rate regions for the ISAC scenario. Finally, numerical results verified that the optimal sensing performance with a superposition of Gaussian and deterministic signals is achieved when the transmit signals are deterministic and the proposed beamforming design significantly enhances the ISAC performance compared with the benchmark schemes.

\appendices

\section{Derivation of the FIM $\mathbf F$}\label{sec:the_derivation_of_the_fim_mathbf_f_}
Let $\dot{\mathbf b}$ denote the partial derivative of $\mathbf b$ w.r.t. $\theta$. Then, we have
\begin{equation}\label{eq:steering_vector_dot}
\dot{\mathbf b}=\frac{\jmath\pi d_\mathrm{a}\cos\theta}{\lambda} \mathbf D_2 \mathbf b,
\end{equation}
with 
\begin{equation}
\mathbf D_2 =\mathrm{diag}\left(-(M_\mathrm{r}-1),-(M_\mathrm{r}-3),\cdots,(M_\mathrm{r}-1)\right).
\end{equation}
Obviously, the derivative of steering vector in \eqref{eq:steering_vector_dot} is orthogonal to the steering vector in \eqref{eq:steering_vector}, i.e.,
\begin{equation}
\mathbf b^H \dot{\mathbf b}  = \dot{\mathbf b}^H \mathbf b = 0.
\end{equation}
Next, the partial derivatives of $\mathbf R_\mathrm{n}$ w.r.t. $\theta$ and $|\alpha|$ are respectively given as
\begin{subequations}
\begin{align}\notag
\frac{\partial \mathbf R_\mathrm{n}}{\partial \theta}&=|\alpha|^2\mathbf I_T \otimes \left( \mathbf a^T \mathbf R_\mathrm{c} \mathbf a^*(\dot{\mathbf b} \mathbf b^H+\mathbf b \dot{\mathbf b}^H)\right),\\
\frac{\partial \mathbf R_\mathrm{n}}{\partial |\alpha|}&= 2|\alpha|\mathbf I_T \otimes \left(\mathbf a^T \mathbf R_\mathrm{c} \mathbf a^*\mathbf b \mathbf b^H\right).
  \end{align}
\end{subequations}
Furthermore, we have
\begin{equation}
\mathbf R_\mathrm{n}^{-1} = \mathbf I_{T}\otimes\frac{1}{\sigma^2}\left(\mathbf I - \frac{|\alpha|^2\mathbf a^T \mathbf R_\mathrm{c} \mathbf a^* \mathbf b \mathbf b^H/\sigma_\mathrm{s}^2}{1+\gamma_\text{ran}(\mathbf R_\mathrm{c})}\right),
\end{equation}
where $\gamma_\text{ran}(\mathbf R_\mathrm{c})= |\alpha|^2\mathbf a^T \mathbf R_\mathrm{c} \mathbf a^*\|\mathbf b\|^2/\sigma_\mathrm{s}^2$.
Finally, based on \eqref{eq:FIM}, each element of $\mathbf F$ is respectively expressed as \eqref{eq:F_1}, \eqref{eq:F_2}, and \eqref{eq:F_3}.

\section{Proof of Proposition~\ref{prop:solution_P3}}\label{sub:proof_of_prop_solution_P3}
Let $\mu_1$, $\mu_2$, and $\mathbf Z\succeq \mathbf 0$ denote the dual variables associated with the constraints \eqref{constr:SINR_random},~ \eqref{constr:semi_random},~\text{and}~\eqref{constr:power_random}, respectively. 
The Lagrangian of problem (P2.1) is
\begin{equation}
\begin{split}
\mathcal L (\mathbf R_\mathrm{c},\mu_2,\mu_1,\mathbf Z)=&~  \mathbf a^T\mathbf R_\mathrm{c}\mathbf a^* -\mu_1 \left(\gamma_0\sigma_\mathrm{c}^2-\mathbf h^H \mathbf R_\mathrm{c}\mathbf h\right)\\
&- \mu_2\left(\mathrm{tr}(\mathbf R_\mathrm{c})- P\right)+ \mathrm{tr}(\mathbf Z \mathbf R_\mathrm{c}).
\end{split}
\end{equation}
The corresponding Karush-Kuhn-Tucker (KKT) conditions are given as \cite{boyd2004convex}
\begin{subequations}
\begin{align}
\frac{\partial \mathcal L (\mathbf R_\mathrm{c},\mu_2,\mu_1,\mathbf Z)}{\partial \mathbf R_\mathrm{c}} &= \mathbf a \mathbf a^H +\mu_1 \mathbf h^* \mathbf h^T-\mu_2 \mathbf I +\mathbf Z^T = \mathbf 0,\\
\mu_1 \left(\gamma_0\sigma_\mathrm{c}^2-\mathbf h^H \mathbf R_\mathrm{c}\mathbf h\right) &=0,\\
\mu_2\left(\mathrm{tr}(\mathbf R_\mathrm{c})- P\right) &= 0,\\
\mathrm{tr}(\mathbf Z \mathbf R_\mathrm{c})&=0,\\
\gamma_0\sigma_\mathrm{c}^2-\mathbf h^H \mathbf R_\mathrm{c}\mathbf h &\le 0,\\
\mathrm{tr}(\mathbf R_\mathrm{c})- P &\le 0,\\
\mathbf R_\mathrm{c} &\succeq \mathbf 0.
\end{align}
\end{subequations}
By solving the KKT conditions, the closed-form solution of problem (P2.1) is 
\begin{equation}
\begin{split}
&\mathbf R_\mathrm{c}^\star =\\
 &\begin{cases} 
P\frac{\mathbf a^*\mathbf a^T}{\|\mathbf a\|^2}, \quad &\text{if~} P|\mathbf h^H\mathbf a^*|^2 \ge M_\mathrm{t} \gamma_0\sigma_\mathrm{c}^2,\\
[\mathbf u_1, \mathbf u_2]
\begin{bmatrix}
\lambda_1 & \lambda_{12}\\
\lambda_{12}^* & \lambda_2
\end{bmatrix}
[\mathbf u_1, \mathbf u_2]^H,\!&\text{otherwise},
\end{cases}
\end{split}
\end{equation}
where 
\begin{equation}
\mathbf u_1=\frac{\mathbf h}{\|\mathbf h\|}, \quad \mathbf u_2 = \frac{\mathbf a^*-\mathbf u_1^H\mathbf a^*\mathbf u_1}{\|\mathbf a^*-\mathbf u_1^H\mathbf a^*\mathbf u_1\|},
\end{equation}
\begin{equation}
\lambda_1= \frac{\gamma_0\sigma_\mathrm{c}^2}{\|\mathbf h\|^2}, \quad \lambda_2 = P- \frac{\gamma_0\sigma_\mathrm{c}^2}{\|\mathbf h\|^2}, \quad \lambda_{12} = \sqrt{\lambda_1\lambda_2}\frac{\mathbf u_1^H\mathbf a^*}{|\mathbf u_1^H\mathbf a^*|}.
\end{equation}

\section{Derivation of the FIM $\hat{\mathbf F}$}\label{sec:the_derivation_of_the_fim_mathbf_hat_f_}
First, the partial derivatives of $\mathbf R_\mathrm{n}$ with respect to $\mathrm{Re}\{\alpha\}$ and $\mathrm{Im}\{\alpha\}$ are respectively given as
\begin{subequations}
\begin{align}
\frac{\partial \mathbf R_\mathrm{n}}{\partial \mathrm{Re}\{\alpha\}}&= 2\mathrm{Re}\{\alpha\}\mathbf a^T \mathbf R_\mathrm{c} \mathbf a^* (\mathbf I_T \otimes \mathbf b \mathbf b^H),\\
\frac{\partial \mathbf R_\mathrm{n}}{\partial \mathrm{Im}\{\alpha\}}&= 2\mathrm{Im}\{\alpha\}\mathbf a^T \mathbf R_\mathrm{c} \mathbf a^* (\mathbf I_T \otimes \mathbf b \mathbf b^H).
  \end{align}
\end{subequations}
Next, the partial derivatives of $\tilde{\mathbf u}_\mathrm{s}$ with respect to $\theta$, $\mathrm{Re}\{\alpha\}$, and $\mathrm{Im}\{\alpha\}$ are respectively given as
\begin{subequations}
	\begin{align}
\frac{\partial \tilde{\mathbf u}_\mathrm{s}}{\partial\theta} &= \left(\mathbf I_T \otimes \alpha \dot{\mathbf b} \mathbf a^T\right) \tilde{\mathbf x}_0,\\
\frac{\partial\tilde{\mathbf u}_\mathrm{s}}{\partial \mathrm{Re}\{\alpha\}} &=  \left(\mathbf I_T \otimes \mathbf b \mathbf a^T\right) \tilde{\mathbf x}_0,\\
\frac{\partial \tilde{\mathbf u}_\mathrm{s}}{\partial \mathrm{Im}\{\alpha\}} &=  \jmath\left(\mathbf I_T \otimes \mathbf b \mathbf a^T\right) \tilde{\mathbf x}_0.
  \end{align}
\end{subequations}
Based on \eqref{eq:FIM_both}, each element of FIM $\hat{\mathbf F}$ is respectively given in \eqref{eq:F_hat_1}, \eqref{eq:F_hat_2}, and \eqref{eq:F_hat_4}.

\ifCLASSOPTIONcaptionsoff
  \newpage
\fi

\bibliographystyle{IEEEtran}
\bibliography{IEEEabrv,mybibfile}

\end{document}